\documentclass[journal]{IEEEtran}
\usepackage{amsmath,amsfonts}
\usepackage{algorithm}
\usepackage{array}
\usepackage{textcomp}
\usepackage{stfloats}
\usepackage{url}
\usepackage{verbatim}
\usepackage{graphicx}
\usepackage{cite}
\usepackage{caption}
\newtheorem{definition}{Definition}
\usepackage{subcaption}
\usepackage{booktabs}
\usepackage{multirow}
\usepackage{amsmath,amsfonts}
\usepackage{algpseudocode}
\usepackage{array}
\usepackage{textcomp}
\usepackage{stfloats}
\usepackage{url}
\usepackage{verbatim}
\usepackage{graphicx}
\usepackage{cite}
\usepackage{caption}
\usepackage{subcaption}
\usepackage{booktabs}
\usepackage{longtable}
\usepackage[table]{xcolor}
\usepackage{paralist}
\usepackage{multirow}
\usepackage{booktabs}

\begin{document}

\title{Scaling Trust in Quantum Federated Learning: A Multi-Protocol Privacy Design}

\author{Dev Gurung, Shiva Raj Pokhrel,~\IEEEmembership{Senior Member,~IEEE,}
\thanks{Authors are with School of IT, Deakin University; email: d.gurung@deakin.edu.au, shiva.pokhrel@deakin.edu.au}
}

\markboth{Journal of \LaTeX\ Class Files,~Vol.~14, No.~8, August~2026}%
{Shell \MakeLowercase{\textit{et al.}}: A Sample Article Using IEEEtran.cls for IEEE Journals}

\maketitle

\begin{abstract}
Quantum Federated Learning (QFL) promises to revolutionize distributed machine learning by combining the computational power of quantum devices with collaborative model training. Yet, privacy of both data and models remains a critical challenge. In this work, we propose a privacy-preserving QFL framework where a network of $n$ quantum devices trains local models and transmits them to a central server under a multi-layered privacy protocol. Our design leverages Singular Value Decomposition (SVD), Quantum Key Distribution (QKD), and Analytic Quantum Gradient Descent (AQGD) to secure data preparation, model sharing, and training stages. Through theoretical analysis and experiments on contemporary quantum platforms and datasets, we demonstrate that the framework robustly safeguards data and model confidentiality while maintaining training efficiency.
\end{abstract}


\section{Introduction}
Machine learning fundamentally relies on extracting patterns from data and applying this knowledge to tasks such as prediction and inference \cite{decristofaroCriticalOverviewPrivacy2021}. 
Models distill information encoded in datasets and store it in learned parameters, typically improving with access to large volumes of data. 
Yet sharing raw data is often infeasible due to privacy, ethical, and regulatory constraints. 
Federated learning mitigates this challenge by enabling distributed training on local devices while ensuring that raw data never leave the device.

Quantum computing has advanced rapidly, with industry leaders such as Google, IBM, and Microsoft pursuing quantum supremacy. 
Given the potential for exponential gains in quantum information processing, machine learning is expected to benefit significantly from quantum implementations \cite{liu2023classical, mitaraiQuantumCircuitLearning2018, cowlessur2025hybrid}.

Privacy risks in machine learning arise from multiple vectors \cite{decristofaroCriticalOverviewPrivacy2021}. 
Adversaries may extract sensitive patterns, reverse engineer models, or exploit memorized information to recover details of the training data. 
Existing countermeasures, including differential privacy, DP-PCA, and t-SNE highlight the essential trade-off between model utility and privacy protection. 
Further threats include membership inference attacks, which attempt to determine whether a specific data instance appears in a model’s training set \cite{liu2025differentially, songSystematicEvaluationPrivacy2021}.

\begin{figure}
    \centering
    \includegraphics[width=0.78\linewidth]{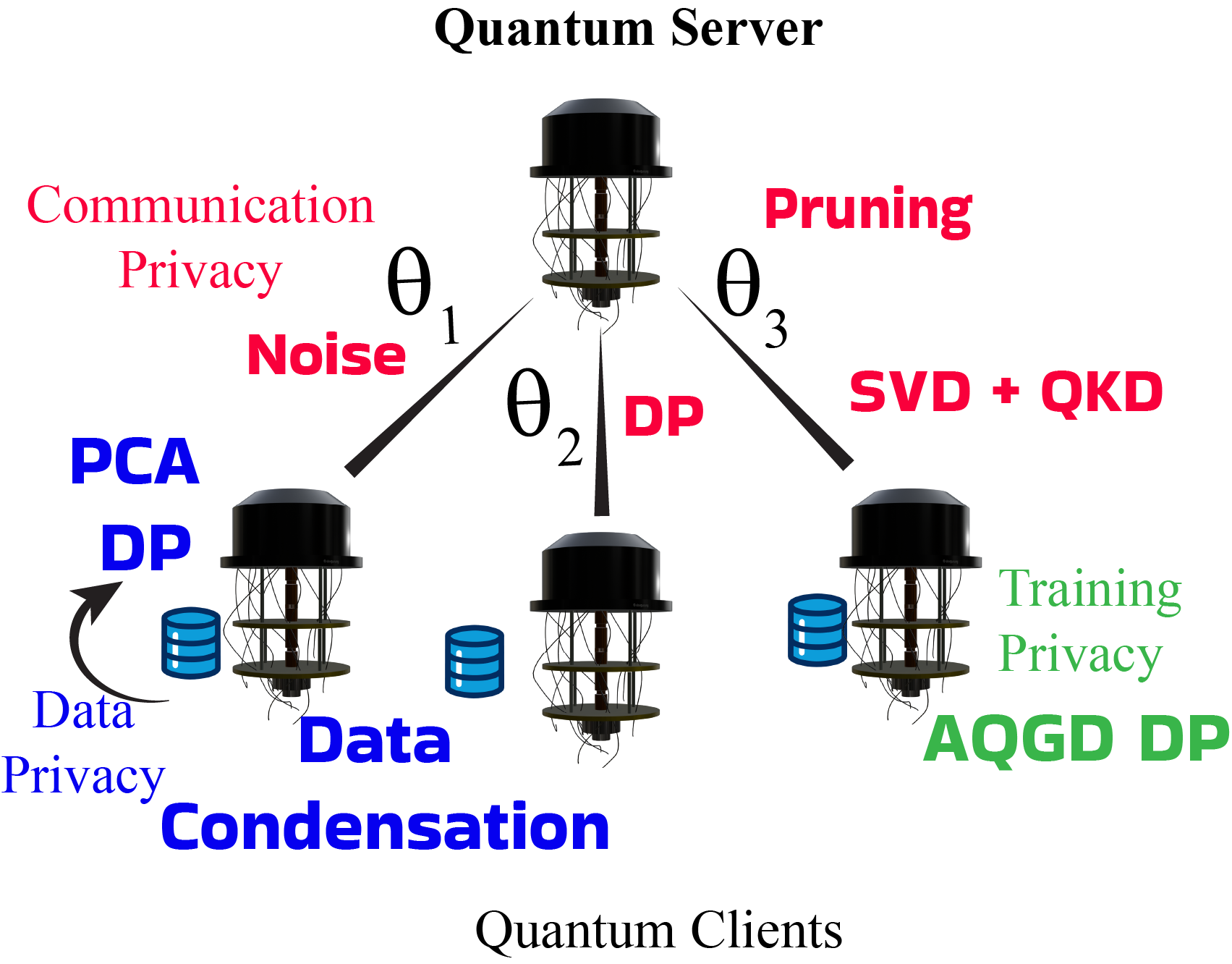}
   \caption{Overview of the proposed privacy-preserving QFL framework. A network of $n$ quantum devices trains local models and transmits them to the server after applying privacy mechanisms across data preparation, model sharing, and training stages. The design integrates Singular Value Decomposition (SVD), Quantum Key Distribution (QKD), and Analytic Quantum Gradient Descent (AQGD) to enable a robust multiprotocol privacy layer.}
    \label{fig:privacyQFL_preview}
\end{figure}

Among privacy-preserving approaches, differential privacy (DP) is widely regarded as the gold standard, offering a rigorous framework and a well-defined threat model for adversarial capabilities \cite{papernotScalablePrivateLearning2018, song2025towards}. 
Traditional methods such as anonymization are vulnerable to re-identification and linkage attacks \cite{DifferentialPrivacyFlowerFramework}. 
DP is a mature research area, and numerous studies have integrated it into iterative training algorithms such as SGD \cite{mcmahanGeneralApproachAdding2019}. 
Most DP-based techniques follow a common principle: compute model updates from training data and then apply a Gaussian DP mechanism to sanitize these updates.

Quantum federated learning (QFL)~\cite{gurung_personalized_2024, pokhrel2024data} is rapidly emerging, with growing research on personalized QFL , federated quantum natural gradient descent \cite{bu_quantum_2021, narottama_federated_2023}, and related methods. 
While privacy preservation has been extensively studied in classical FL \cite{jiLessMoreRevisiting}, comparable exploration within QFL remains limited.

A clear understanding of \emph{what}, \emph{when}, and \emph{how} attacks arise in FL systems is essential. 
The \emph{what} identifies the attack targets: data privacy, the learned model, and communication channels. 
The \emph{when} refers to the adversarial stages, including training, inference, and communication \cite{decristofaroCriticalOverviewPrivacy2021}. 
The \emph{how} captures concrete attack mechanisms such as statistical disclosure, model inversion, and class-representative inference. 
In FL, these manifest through local model poisoning, distributed backdoor attacks, and related adversarial strategies \cite{jereTaxonomyAttacksFederated2021}.

Given the significant gap in privacy-preserving methods for QFL, the main contributions of this work are:

\begin{enumerate}
    \item We introduce a privacy-preserving QFL framework and identify privacy bottlenecks that arise uniquely in quantum settings. The distinctive characteristics of QFL demand a privacy perspective fundamentally different from classical FL.
    
    \item We propose a suite of new protocols and demonstrate how existing techniques can be combined to enhance quantum privacy across multiple layers, including data processing, training, and model-parameter exchange. These protocols deliver structured, multi-layered privacy protection tailored to QFL.
    
    \item We provide extensive theoretical and experimental evaluations, detailing our implementations, encountered challenges, and a rigorous assessment of the practicality and suitability of the proposed methods within QFL systems.
\end{enumerate}

\section{Background}
\subsection{Differential Privacy}
DP is an approach for more secure and private distributed machine learning.
In terms of DP for the dataset, the objective is to maintain the identity of the data owners when their data are being used for machine learning.
Two general approaches are local DP where noise is added to raw data before sending them to the server and global DP where noise is added at the global level after
all data from clients are combined \cite{introToDPPyDP}.

In the real world, DP is used by companies such as Uber (elastic sensitivity to limit the access to the traffic query and driver revenue by staff), Apple (local DP to randomize raw data to analyze user behavior and improve user experience) and Google (e.g., Google keyboard with FL) \cite{introToDPPyDP}.

For a simple example of DP, let us suppose that there are two datasets D and D'. 
They are identical except for a single record (which could be one row).
With differential privacy, we can assure that for any analysis of the datasets such as calculation of average value, the result will be similar \cite{DifferentialPrivacyFlowerFramework}. 
One of the commonly used approaches to achieve this is to add noise to the output of the analysis to mask the contribution of each individual in the data while still preserving the result of the analysis~\cite{song2025towards}.

\begin{definition}[(\(\varepsilon, \delta\))-Differential Privacy]
   A randomized algorithm $M$ provides $(\epsilon, \delta)$-differential privacy if for any two neighboring databases, $x$ and $y$ which differ in a single record and for all the possible outputs $S \subseteq Range(M)$, $x,y \in N$ such that $\|x-y\|_1 \leq 1$ \cite{dworkAlgorithmicFoundationsDifferential2013}:
\begin{equation}
    P[M(x) \in S] \leq e^\epsilon P[M(y) \in S] + \delta
\end{equation}
 where, $\epsilon$ is privacy budget term for metric of privacy loss which controls the privacy-utility trade-off and $\delta$ parameter accounts for small probability on which the upper bound $\epsilon$ does not hold.
 The lower value for $\epsilon$ indicates the higher levels of privacy with trade-off with reduced utility as well.
\end{definition}
Various mechanisms are used to achieve differential privacy.
One of the classic Gaussian mechanisms is proposed by Dwork et al. \cite{dworkAlgorithmicFoundationsDifferential2013}, which is essential to achieve $(\epsilon, \delta)-$DP \cite{jiLessMoreRevisiting}.

Differential private stochastic gradient descent is a differential private version of the mini-batch stochastic optimization process.
The idea behind this is that we can access the loss gradient with respect to each parameter (parameter gradient) of our model.

\subsection{Data Condensation}
Data condensation addresses two challenges, the expensive training cost and poor generalization performance due to data generators that are used to produce differentially private data for model training \cite{dongPrivacyFreeHow2022}.

Consider a large dataset $D = \{(x_i, y_i)\}_{i=1}^{|D|}$ with inputs $x_i \in \mathbb{R}^d$ and labels $y_i \in \{0, \dots, C-1\}$. 
Data condensation seeks to construct a much smaller synthetic dataset $S = \{(s_i, y_i)\}_{i=1}^{|S|}$ with $|S| \ll |D|$ such that training a model $\phi_\theta$ in $S$ by minimizing $L_S(\theta) = \frac{1}{|S|} \sum_{(s,y) \in S} \ell(\phi_\theta(s), y)$ yields generalization performance comparable to training in $D$ (by minimizing $L_D(\theta)$).

The parameter matching framework expresses this goal as
\[
\min_S \mathbb{E}_{\theta_0 \sim P_{\theta_0}} \big[ D(\theta_S(\theta_0), \theta_D(\theta_0)) \big],
\]
where, $\theta_S = \arg\min_\theta L_S(\theta)$ and $\theta_D = \arg\min_\theta L_D(\theta)$. 
Here, $D(\cdot, \cdot)$ quantifies the similarity between the set of parameters, encouraging $\theta_S \approx \theta_D$ (for the models trained on $S$ and $D$, respectively) so that the performance remains stable across random initializations $\theta_0$ \cite{dongPrivacyFreeHow2022}.

\subsection{Quantum Key Distribution}
Quantum Key Distribution (QKD) enables the exchange of random bits over a communication channel without the need for any pre-shared secret information \cite{bennettQuantumCryptographyPublic2014}. 
The parties communicate using a conventional, classical channel that may be subject to passive eavesdropping.
Based on whether any disturbance is detected during transmission, they then decide whether to accept or discard the resulting secret key.
QKD is based on fundamental concepts of quantum mechanics, which are Heisenberg's uncertainty principle, the no-cloning Theorem, and quantum entanglement.
In brief, Heisenberg's uncertainty principle states that a pair of physical properties such as position and momentum cannot be measured simultaneously. That means we cannot know both the position and the speed of a particle such as a photon or electron with high accuracy, which can be mathematically presented as 
\[
\Delta x \Delta p \geq \frac{h}{4 \pi}
\]
where, $h$ is Planck's constant, $\Delta x$ and $\Delta p$ is uncertainty in position and momentum, respectively.
Quantum cryptography utilizes the polarization of photons on different bases as conjugate properties.
With no-cloning theorem, an eavesdropper cannot make a perfect copy of unknown quantum state.
With quantum entanglement, two quantum particles can be entangled with each other.

\subsection{Analytic Quantum Gradient Descent (AQGD)}
The AQGD optimizer is a first-order optimization algorithm designed specifically for variational quantum algorithms employing parameterized Pauli rotation gates. 
It performs a momentum-accelerated gradient descent using analytic
 gradients (based on parameter-shift) computed directly on quantum hardware with only $2n+1$ circuit evaluations per step, where $n$ is the number of parameters. 
Key features include epoch-based scheduling of the learning rate $\eta$, momentum coefficient $\in [0,1)$, 
and flexible convergence criteria that combine: 
(i) tolerance on a moving-window average of the objective function (tol and averaging), and 
(ii) tolerance on the $\infty$-norm of parameter updates (param\_tol). 
This makes AQGD particularly effective and hardware-efficient for training quantum circuits, as introduced by Mitarai et al. \cite{mitaraiQuantumCircuitLearning2018} and extended with analytic gradient techniques in Schuld et al.\cite{schuldEvaluatingAnalyticGradients2019}.

\subsection{Singular Value Decomposition}
Singular value decomposition (SVD) is a powerful linear algebra method for matrix factorization, notable for its numerical robustness and guaranteed existence \cite{brezinaSupramolecularComplexesOxoporphyrinogens2024}.
For a matrix A, its SVD represents A as a product of three matrices, $A = UDV^T$, where the columns of $U$ and $V$ form orthonormal sets, and $D$ is a diagonal matrix whose diagonal entries are positive real numbers.
The rectangular matrix D is a diagonal matrix whose non-zero entries are called singular values.  
SVD serves as a technique for dimensionality reduction and has a strong connection to principal component analysis (PCA).  
These approaches are applied in a wide range of tasks, including identifying patterns in data and developing face recognition algorithms.

\subsection{Literature}
By default, FL ensures data privacy by keeping data on local devices \cite{mcmahan2017communication}. However, achieving a robust and trustworthy FL requires additional defenses. For instance, Robust-DPFL \cite{qiRobustnessDifferentiallyPrivate2024a} distinguishes poisoned and clean local gradients to robustly update the global model. Extensions like BLT-DP-FTRL \cite{mcmahanHasslefreeAlgorithmStrong2024a} maintain the ease-of-use advantages of tree aggregation while supporting multiplication scenarios.  

Data condensation techniques \cite{zhaoDATASETCONDENSATIONGRADIENT2021} improve training efficiency by generating synthetic datasets, further enhanced by DP integration in PPFL-DC \cite{zhangPrivacyPreservingFederatedLearning2025} to protect model weights and reduce communication costs. Similar approaches in vertical FL \cite{gaoSecureDatasetCondensation2024} focus on efficiency and privacy via synthetic data, while PQSF \cite{zhangPQSFPostquantumSecure2024a} uses double masking and secret sharing for post-quantum secure FL.  

Other classical FL privacy methods include RCFL \cite{zhouRobustFederatedLearning2025}, PriVeriFL-A \cite{wangPriVeriFLPrivacyPreservingAggregationVerifiable2025}, and DConBe-based decentralized schemes \cite{zhangPrivacyPreservingFederatedLearning2025}, which combine DP, homomorphic encryption, and aggregation strategies. Further approaches address participation-dependent privacy \cite{qinParticipationDependentPrivacyPreservation2025}, poisoning-resilient FL \cite{zhaoEfficientPrivacyPreservingFederated2024}, and cross-silo privacy enhancement \cite{liEfficientPrivacyEnhancedFederated2024}.  

Overall, most classical FL privacy work centers on DP in classical settings. In contrast, our approach operates in a QFL with quantum clients, implementing multi-layered privacy mechanisms beyond DP~\cite{song2025towards} to achieve a comprehensive privacy-preserving QFL framework.

\begin{figure*}
    \centering
    \includegraphics[width=\linewidth]{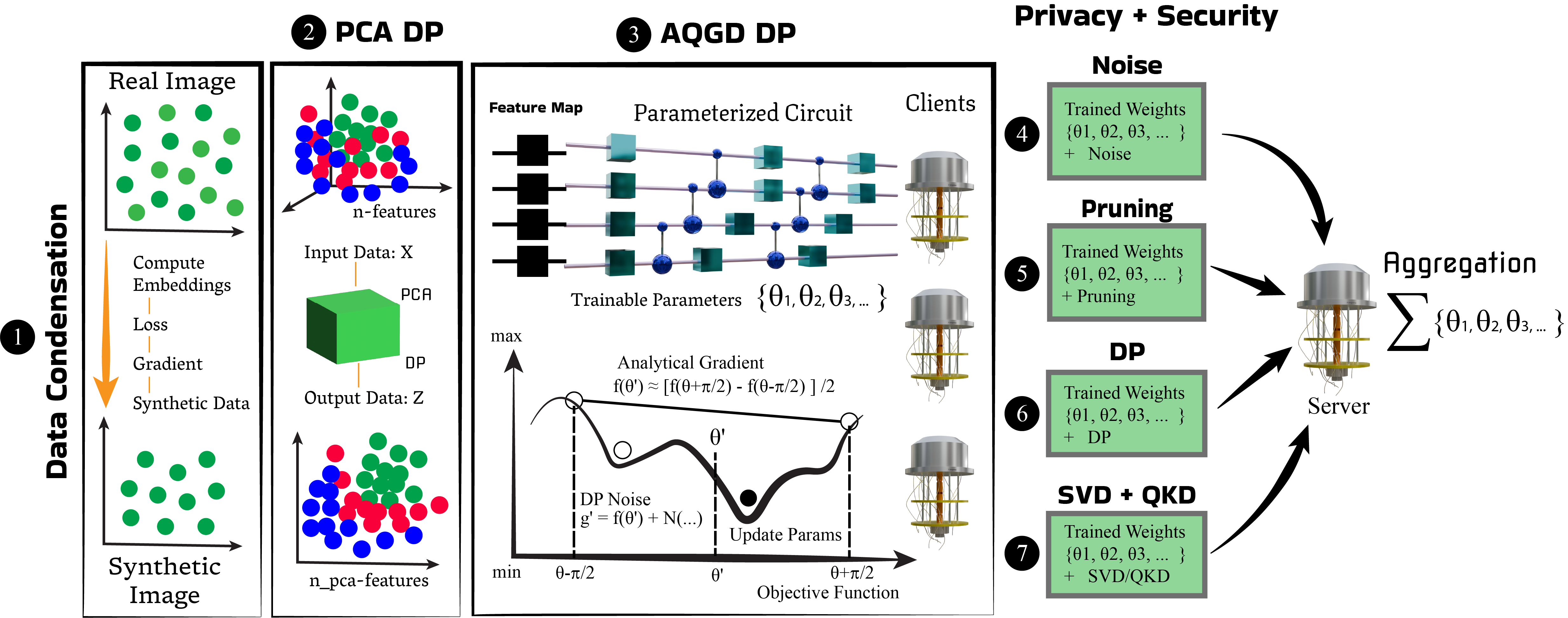}
    \caption{Privacy-Preserving QFL Framework: Various protocols are implemented to provide further privacy to the QFL framework that include privacy during training, data privacy and communication privacy; \textcircled{1} Data condensation, \textcircled{2} PCA DP, \textcircled{3} AQGD DP, \textcircled{4} Noise, \textcircled{5} Pruning, \textcircled{6} DP, \textcircled{7} SVD + QKD}
    \label{fig:privacyQFL_framework}
\end{figure*}

\section{Proposed Framework}
First, we identify of privacy points for privacy bottlenecks in the aspect of QFL.
There can be various privacy bottlenecks that come with QFL.
\begin{inparaenum}[(a)]
 \item Data: The first privacy issue concerns the data itself.
    \item Model parameters: The second type of privacy risk arises from the model parameters, which can be exposed to unauthorized parties who might access, manipulate, or tamper with them.
    \item Optimization: A persistent privacy concern exists during optimization via gradient descent, especially regarding limiting the influence of individual data samples. This is essential for protecting sensitive information and preventing the model from revealing personally identifiable data.
    \item Data Processing: Strengthening privacy is also advantageous for data processing tasks, such as those involving principal component analysis.
\end{inparaenum}

\subsection{System Design}
The main building blocks of the proposed framework are the following.
\begin{inparaenum}[(a)]
    \item Noise: Injecting noise into local models, the server obtains perturbed updates, which safeguards the client's data against inference attacks.
    \item PCA-DP: Applying PCA-DP to preprocess the datasets.
    \item Privacy aspects: In terms of additional privacy considerations, we explore several alternative methods, including data condensation, SVD combined with QKD-based encryption and decryption, as well as model pruning techniques designed to keep the model parameters confidential.
 \item Pruning: We introduce a pruning strategy that transmits only a subset of the model’s parameters.
    \item SVG: By incorporating SVD-based integration, we ensure the privacy of the model parameters.
    \item Condensation: Using data condensation, we derive a compact synthetic dataset that preserves model performance while substantially reducing the data size.
    \item Optimizer DP: Incorporating privacy into the optimization procedure.
\end{inparaenum}
The details of the modules are provided below.

\begin{algorithm}
\caption{DP Noise Addition for Parameters}
\label{alg:dp-noise}
\begin{algorithmic}[1]
\State \textbf{Input}: Parameters $\theta \in \mathbb{R}^n$, privacy budget $\epsilon$, sensitivity $s$, mechanism $\in \{\text{``Laplace''}, \text{``gaussian''}\}$, $\delta$ (for Gaussian)
\State \textbf{Output}: Noisy parameters $\tilde{\theta}$
\If{$\text{mechanism} = \text{``Laplace''}$}
    \State $\text{scale} \gets s / \epsilon$ 
    \State $\text{noise} \gets \text{Laplace}(0, \text{scale}, \text{size}=\theta.\text{shape})$ 
\Else
    \State $\text{scale} \gets s / \epsilon$
    \State $\sigma \gets \text{scale} \cdot \sqrt{2 \ln(1.25 / \delta)}$ 
    \State $\text{noise} \gets \mathcal{N}(0, \sigma^2, \text{size}=\theta.\text{shape})$ 
\EndIf
\State $\tilde{\theta} \gets \theta + \text{noise}$ 
\State \textbf{Return}: $\tilde{\theta}$
\end{algorithmic}
\end{algorithm}

\textbf{Noise.}
Adding noise to the parameters is a useful approach that can keep the parameters private as in Algorithm \ref{alg:dp-noise}.
However, the trade-off is that noise can introduce a performance bottleneck.

\begin{algorithm}
\caption{DP-PCA Preprocessing for Dataset}
\label{alg:dp-pca}
\begin{algorithmic}[1]
\State \textbf{Input}: Dataset features $X \in \mathbb{R}^{m \times d}$, labels $y \in \{0, 1, 2\}^m$, privacy budget $\epsilon$
\State \textbf{Output}: Train/test splits $\hat{X}_{\text{train}}, \hat{X}_{\text{test}}, y_{\text{train}}, y_{\text{test}}$
    \State Load $X \gets \text{dataset.data}$, $y \gets \text{dataset.target}$ 
        \State $\text{bounds} \gets (\min(X, \text{axis}=0), \max(X, \text{axis}=0))$ 
        \State $\text{data\_norm} \gets \max(\|X_i\|_2, i=1,\dots,m)$ 
        \State $\text{pca} \gets \text{PCA\_DP}(n_{\text{components}}, \epsilon, \text{bounds}, \text{data\_norm})$ \cite{imtiazSymmetricMatrixPerturbation2016}
        \State $\hat{X} \gets \text{pca.fit\_transform}(X)$ 
\State \textbf{Return}: $\hat{X}_{\text{train}}, \hat{X}_{\text{test}}, y_{\text{train}}, y_{\text{test}}$
\end{algorithmic}
\end{algorithm}

\textbf{DP-PCA.}
Algorithm \ref{alg:dp-pca} shows the implementation details for DP-PCA.
PCA is one of the popular tools used for dimension reduction.
This is crucial because many times it is hard to work computationally with data of higher dimension.
DP-PCA adds a differential privacy guarantee on top of PCA \cite{imtiazSymmetricMatrixPerturbation2016}. 
From \cite{imtiazSymmetricMatrixPerturbation2016}, the DP PCA procedure consists of the following steps. We begin by specifying the parameters: the privacy budget $\epsilon$, the data norm (i.e., the maximum permitted L2 norm of each row $x_i$ in the dataset), and other bounds that are required for the privacy budget and sensitivity assumptions.
For $m$ features, $n$ samples, input data $X$, individual data point $x_i$, $k$ retained principal components, sample mean vector $\vec{X}$, privacy budget $\epsilon$, and $\delta$ denoting the failure probability of the approximate DP guarantee, we calculate the mean under differential privacy as
\[
\vec{\mu} = mean (X) + noise = \frac{1}{n}\sum_{i=1}^n x_i + \eta, \eta \sim \mathcal{N}(0, \sigma^2_\mu, I_m)
\]
which perturbs the mean so that centering does not reveal information, while also using half of the privacy budget $\epsilon$.

\textbf{DP-AQGD.}
As presented in Algorithm \ref{alg:dp-aqgd}, We aim to bound how much information the model parameters contain about the data. This bound will only capture information that an adversary could realistically extract.
Differential privacy is a widely used method in privacy-preserving frameworks, offering an $(\epsilon, \delta)$-privacy guarantee that limits the impact of any individual training sample on the AQGD training process \cite{abadiDeepLearningDifferential2016}.

\textbf{QKD.}
Quantum key distribution is a quantum approach in which we share the encryption key between the sender and the receiver that is also quantum secure.
QKD in this work is applied to the sigma part of the model parameters, that is, a small part of the parameters, as encrypting whole model parameters which are big might be redundant and unnecessary. 
Encryption can be one-pad based or based on classical cryptography. 

\begin{figure}
    \centering
    \includegraphics[width=0.9\linewidth]{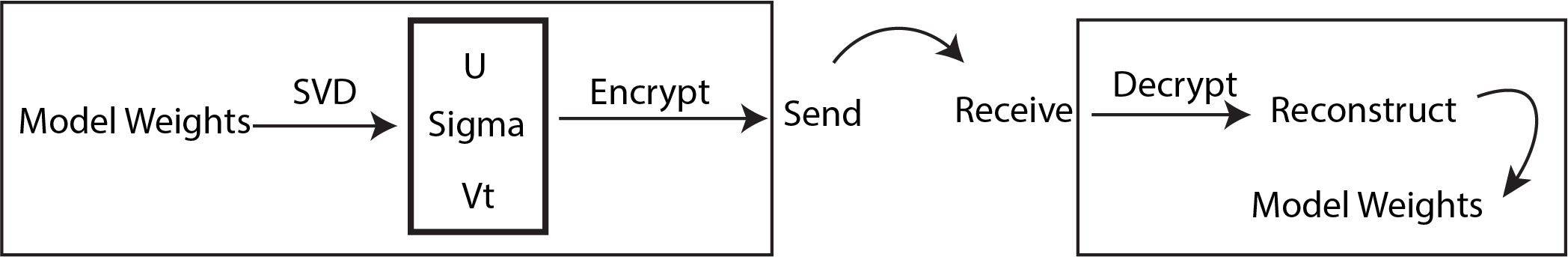}
    \caption{Implementation of SVD based privacy secure QFL framework.}
    \label{fig:svd_encryption}
\end{figure}

\textbf{SVD.}
In this approach (Algorithm \ref{alg:svd_qkd}), as shown in Figure \ref{fig:svd_encryption}, we first use the SVD method to divide the model weights into $U$, $Sigma$, and $Vt$. 
We only encrypt $Sigma$ and send $U$, encrypted $Sigma$ and $Vt$ to the server for averaging.
The server decrypts the encrypted $Sigma$ and reconstructs the original weight matrix from $U$, $Vt$ and the decrypted $Sigma$.
This will give the actual model weights of the particular local device.
This approach guarantees that an eavesdropper can never fully reconstruct the model weights. 
In addition, it reduces the overhead of encryption and decryption, because the sigma component of the weight matrix is significantly smaller than the complete set of model weights.
This preserves the integrity of the model weights while reducing encryption overhead, thereby improving both their performance and privacy.

\begin{figure}
    \centering
    \includegraphics[width=\linewidth]{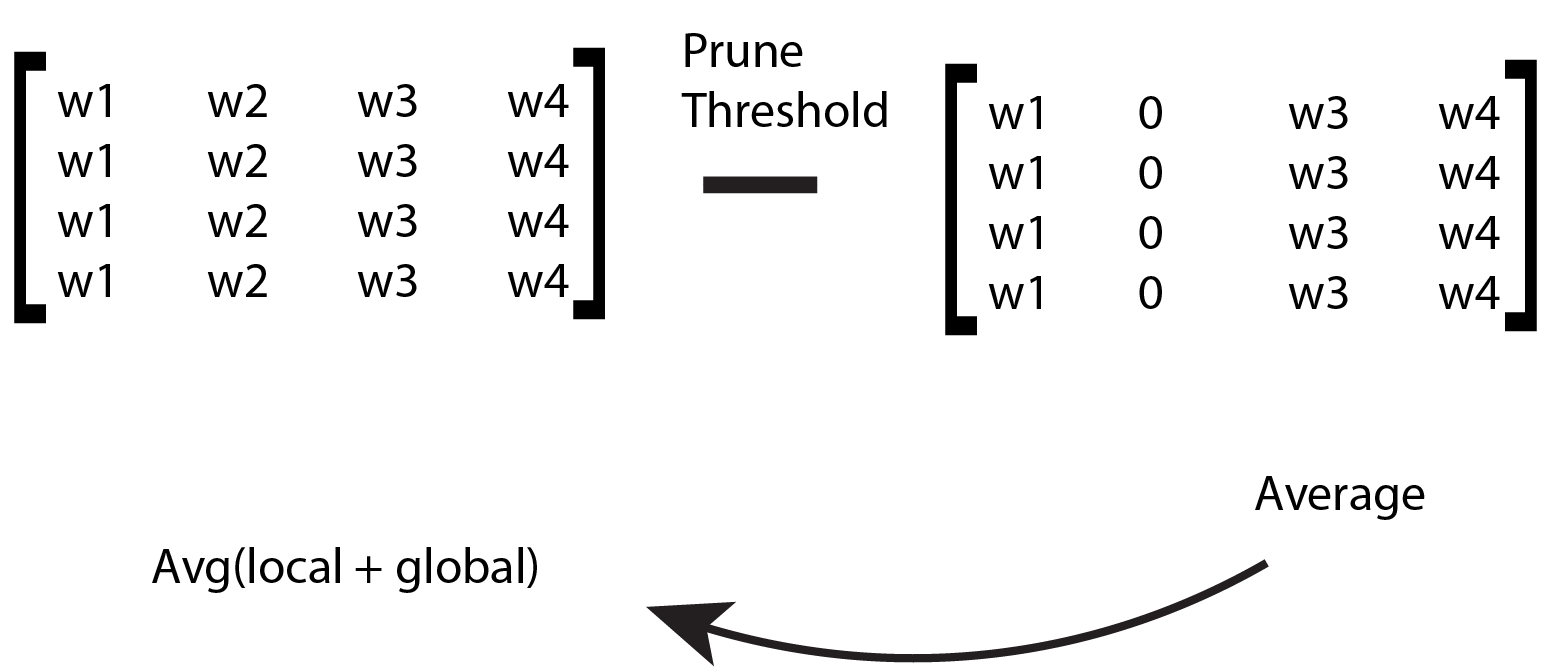}
    \caption{Model pruning approach for QFL privacy}
    \label{fig:pruning}
\end{figure}

\textbf{Pruning.}
In this method (Algorithm \ref{alg:qfl-pruning-combined}), illustrated in Figure \ref{fig:pruning}, we share only a subset of the model weights, specifically either a later segment or the initial portion of the model weight array.
This approach also benefits communication, as an eavesdropper cannot reliably infer information about the dataset from the transmitted model parameters, which constitute only a partial set of the full model weights.
Another facet of this method is that the server may only need depending on additional customization to aggregate and average a subset of the model weights collected from all devices, rather than the entire parameter set.
Any subsequent validation or training on the server side can likewise be carried out using a smaller model architecture, if such an implementation is in place.
After averaging, the resulting global model is more compact; therefore, when transmitted back to the devices for further local training, 
the missing portion can be restored from the previous model of the device and only the overlapping parts are re-averaged with the global weights received.

\textbf{Data Condensation.}
With the data condensation technique, we generate a small synthetic dataset that captures the essential characteristics of a larger dataset to train machine learning models \cite{zhaoDatasetCondensationDistribution2023}.

\textbf{Default Privacy.}
In QFL, the data features are represented as quantum states during training. 
This inherently offers a certain level of protection against reconstructing the original data from the model parameters. 
Consequently, QFL provides stronger privacy guarantee compared to classical FL systems.
However, examining and quantifying this aspect lies beyond the scope of the present work and is therefore left for future research.

\section{Theoretical Analysis}
\subsection{Noise}
In this approach, we add noise to the model parameters to achieve differential privacy.
Consider the device model parameters \(\theta \in \mathbb{R}^d\) as the tunable weight tensor of a variational quantum circuit, consisting of \(d\) elements,
    \( \epsilon \) be the privacy budget,
    \( \Delta \) be the sensitivity of the \( L_1 \) norm of the weights.
    The mechanism can be methods like the Laplace mechanism.

The noise addition process involves scale calculation where the scale parameter \( b \) for the Laplace distribution is computed as
    \[
    b = \frac{\Delta}{\epsilon} 
    \]
    Whereas, in terms of 
    noise sampling, for each component \( i = \{1, 2, \dots, d \}\), an independent noise term \( \eta_i \) is sampled as,
    \[
    \eta_i \sim \text{Laplace}(0, b)
    \]
    The probability density function is obtained as
    \[
    f(x; 0, b) = \frac{1}{2} \exp(-|x|)
    \]
    The noise vector is \( \eta = (\eta_1, \eta_2, \dots, \eta_d) \in \mathbb{R}^d \).
    Finally, the differentially private parameters are
    \[
    \tilde{\theta} = \theta + \eta, \quad \tilde{\theta}_i = \theta_i + \eta_i, \quad \eta_i \sim \text{Laplace}(0, b)
    \]
\subsection{PCA-DP}
PCA is applied to a dataset with differential privacy.
Let \( X \in \mathbb{R}^{n \times m} \) be the input dataset with \( n \) samples and \( m \) features and 
$k$ number of principal components to retain.
With standard PCA, we first compute the covariance matrix of the centered data as
    \[
    \Sigma = \frac{1}{n} (X - \bar{X})^\top (X - \bar{X})
    \]
    where \( \bar{X} \) is the mean of the data in the samples.
Then eigenvalue decomposition follows by computing the top \( k \) eigenvectors \( U_k \in \mathbb{R}^{m \times k} \) of \( \Sigma \), corresponding to the \( k \) largest eigenvalues.
    Finally, transformation is performed by projecting the centered data onto the \( k \)-dimensional subspace as
    \[
    X_{\text{pca}} = (X - \bar{X}) U_k \in \mathbb{R}^{n \times k}
    \]

However,
a differentially private PCA is applied with parameters such as
\( k\) the number of components, \( \epsilon\) the privacy budget, \( \text{bounds}\), assuming that each feature of \( X \) is scaled to \( [0, 1] \) and the \( L_2 \) norm bound for each data point.
The PCA-DP algorithm (e.g., based on differentially private covariance matrix estimation) proceeds by first 
data preprocessing by assuming each data point \( x_i \in \mathbb{R}^m \) (row of \( X \)) is clipped or scaled such that \( \|x_i\|_2 \), and each feature is in \( [0, 1] \).
Then a noisy covariance matrix is computed to ensure differential privacy as
\[
\tilde{\Sigma} = \frac{1}{n} (X - \bar{X})^\top (X - \bar{X}) + N
\]
where, \( N \in \mathbb{R}^{m \times m} \) is a symmetric noise matrix, typically with entries sampled from mechanisms such as the Gaussian or Laplace distribution. 

\begin{algorithm}
\caption{DP-AQGD Optimizer}
\label{alg:dp-aqgd}
\begin{algorithmic}[1]
\State \textbf{Input}: Parameters $\theta \in \mathbb{R}^n$, objective function $f$, max iterations $maxIter$, learning rate $\eta$, tolerance $tol$, momentum $\mu$, parameter tolerance $paramTol$, averaging $Averaging$, privacy budget $\epsilon$, failure probability $\delta$, sensitivity $s$
\State \textbf{Output}: Private parameters $\theta^*$, objective value $f(\theta^*)$
\State $\theta_0 \gets \theta$, $\text{eval\_count} \gets 0$
\For{$t = 1$ to $\text{maxiter}$}
    \State $n \leftarrow $ Length of Parameters
    \State $\Theta \gets$ Perturbed parameters via parameter shift rule.
    \State $\text{values} \gets f(\Theta)$ 
    \State $\text{eval\_count} \gets \text{eval\_count} + 2n + 1$
    \State $f(\theta_t) \gets \text{values}[0]$ 
    \State $g_t \gets \frac{1}{2} (\text{values}[1:n+1] - \text{values}[n+1:2n+1])$ 
    \State $\sigma \gets s \cdot \sqrt{2 \ln(1.25 / \delta)} / \epsilon$ 
    \State Compute noise $\eta = (0, \sigma, gradientSize)$
    \State Noisy Gradient = $g_t + \eta$
    \State Update $\theta_t$
\EndFor
\State \textbf{Return}: $\theta_{t+1}$, $f(\theta_{t+1})$
\end{algorithmic}
\end{algorithm}

\subsection{AQSD DP}
In this work, we extend the AQGD optimizer from Qiskit's machine learning library, incorporating Gaussian noise to achieve $(\epsilon, \delta)$-differential privacy. 
In the following, we formalize the framework of the optimizer.

The DP\_AQGD is initialized with parameters for optimization and differential privacy such as
 $\text{maxiter} \in \mathbb{Z}^+$: maximum number of iterations (gradient steps),
 $\eta \in \mathbb{R}^+$: learning rate for gradient updates,
   $\text{tol} \in \mathbb{R}^+$: tolerance for change in the windowed average of objective function values,
   $\text{momentum} \in [0, 1)$: Momentum coefficient for biasing updates toward previous gradients,
   $\text{param\_tol} \in \mathbb{R}^+$: tolerance for change in the L2 norm of parameters,
   $\text{averaging} \in \mathbb{Z}^+$: window size to average objective function values,
   $\epsilon \in \mathbb{R}^+$: privacy budget controlling the privacy-utility trade-off in $(\epsilon, \delta)$-DP, $\delta \in (0, 1)$: probability of privacy failure in $(\epsilon, \delta)$-DP and $\text{sensitivity} \in \mathbb{R}^+$: maximum L2 norm of the gradient ($\Delta_2 f$).

The optimizer operates on a parameter vector $\theta \in \mathbb{R}^n$, where $n$ is the number of parameters, to minimize a differentiable objective function $f: \mathbb{R}^n \to \mathbb{R}$ using gradient descent with momentum and differential privacy.
This main function is to compute the objective function value and its gradient with Gaussian noise for $(\epsilon, \delta)$-differential privacy that takes the input parameters as
\begin{inparaenum}
    \item $\theta \in \mathbb{R}^n$, which is current parameter vector and 
    \item $f: \mathbb{R}^n \to \mathbb{R}$ which is objective function.
\end{inparaenum}

Among the steps involved, the first step involves parameter perturbations, where $f$ is evaluated at $2n + 1$ points to calculate the gradient analytically with the
current parameters $\theta$ and perturbed parameters $\theta + \frac{\pi}{2} e_i$ and $\theta - \frac{\pi}{2} e_i$ for each basis vector $e_i \in \mathbb{R}^n$, where $e_i$ is the $i$-th standard basis vector.
These points form a matrix

\[
\begin{split}
\Theta^T = \begin{bmatrix}
    \theta & \theta + \frac{\pi}{2} e_1 & \dots & \theta + \frac{\pi}{2} e_n & \dots \\
\end{bmatrix} \\
\begin{bmatrix}
    \dots & \theta - \frac{\pi}{2} e_1 & \dots & \theta - \frac{\pi}{2} e_n
\end{bmatrix}
\in \mathbb{R}^{n \times (2n+1)}
\end{split}
\]

The objective function is evaluated as
\[
v^T = \begin{bmatrix}
    f(\theta) & f\left(\theta + \frac{\pi}{2} e_1\right) & \dots & f\left(\theta - \frac{\pi}{2} e_n\right)
\end{bmatrix} \in \mathbb{R}^{1 \times (2n+1)}
\]

The evaluation count is incremented as 
    $\text{eval\_count} \gets \text{eval\_count} + 2n + 1.$
The objective value is:
    \[
    f(\theta) = v[0]
    \]
For gradient computation,   
the gradient is computed using a finite-difference-like (parameter shift rule) method
\[
\nabla f(\theta)_i = \frac{f\left(\theta + \frac{\pi}{2} e_i\right) - f\left(\theta - \frac{\pi}{2} e_i\right)}{2}
\]
For $i = \{1, \dots, n$\}, the gradient vector is:
\[
\begin{split}
\nabla f(\theta)^T = \begin{bmatrix}
    \frac{v[1] - v[n+1]}{2} & \frac{v[2] - v[n+2]}{2} & \dots & \dots \\
\end{bmatrix} \\
\begin{bmatrix}
    \dots & \frac{v[i] - v[n+i]}{2} & \dots & \frac{v[n] - v[2n]}{2}
\end{bmatrix}
\in \mathbb{R}^{1 \times n}
\end{split}
\]
Now for the privacy mechanism 
Gaussian noise is added to ensure $(\epsilon, \delta)$-DP with 
L2 sensitivity of the gradient $\Delta_2 f = \text{sensitivity}$, defined as
\[
\Delta_2 f = \sup_{D, D'} \|\nabla f(\theta; D) - \nabla f(\theta; D')\|_2
\]
where $D, D'$ are neighboring datasets (differing by one record).
The standard deviation of the Gaussian noise is
\[
\sigma = \frac{\Delta_2 f \sqrt{2 \ln \left( \frac{1.25}{\delta} \right)}}{\epsilon}
\]
Independent Gaussian noise is sampled as
\[
z \sim \mathcal{N}(0, \sigma^2 I_n)
\]
where $z \in \mathbb{R}^n$ and $I_n$ is the $n \times n$ identity matrix.
The final noisy gradient obtained is 
\[
\tilde{\nabla} f(\theta) = \nabla f(\theta) + z
\]
with results obtained as
\[
(f(\theta), \tilde{\nabla} f(\theta)).
\]

The privacy mechanism ensures $(\epsilon, \delta)$-differential privacy. For a function with L2 sensitivity $\Delta_2 f$, adding noise $\mathcal{N}(0, \sigma^2 I_n)$ with
\[
\sigma \geq \frac{\Delta_2 f \sqrt{2 \ln \left( \frac{1.25}{\delta} \right)}}{\epsilon}
\]
guarantee
$\Pr[M(\theta; D) \in S] \leq e^\epsilon \Pr[M(\theta; D') \in S] + \delta$
for any measurable set $S$, where $M(\theta; D) = \tilde{\nabla} f(\theta)$ is the noisy gradient.
The parameter update uses gradient descent with momentum
$m_t = \text{momentum} \cdot m_{t-1} + (1 - \text{momentum}) \cdot \tilde{\nabla} f(\theta_t)$
and 
$\theta_{t+1} = \theta_t - \eta \cdot m_t$
where, $m_t \in \mathbb{R}^n$ is the momentum vector at iteration $t$. 

\begin{algorithm}
\caption{QFL with Pruning and Averaged Initial Weights}
\label{alg:qfl-pruning-combined}
\begin{algorithmic}[1]
\State \textbf{Input}: $K$ devices with VQC weights $\theta_k \in \mathbb{R}^n$, server VQC, $T$ rounds, pruning threshold $\tau = 0.5$, average initial flag $\text{avg\_initial} \in \{True, False\}$
\State \textbf{Output}: Aggregated weights $\bar{\theta}^{(T-1)}$
\For{each round $t = \{0, 1, \dots, T-1\}$}
    \State $\bar{\theta}^{(t-1)} \gets \text{null}$ (for $t = 0$)
    \State $\text{total\_weights} \gets []$
    \For{each device $k = 1, 2, \dots, K$}
        \State $t_k \gets t$
        \If{$t > 0$}
            \If{$\text{avg\_initial}$}
                \State $\theta_k^{(t)} \gets \frac{\bar{\theta}^{(t-1)} + \theta_k^{(t)}}{2}$ 
            \Else
                \State $\theta_k^{(t)} \gets \bar{\theta}^{(t-1)}$ 
            \EndIf
        \EndIf
        \State $\theta_k^{(t)} \gets \text{Train}(\theta_k^{(t)}, f_k)$ 
        \Comment{Local VQC training}
        \State $\tilde{\theta}_k^{(t)} \gets \theta_k^{(t)}$ 
        \State $\tilde{\theta}_k^{(t)}[|\tilde{\theta}_k^{(t)}| < \tau] \gets 0$ 
        \State $\text{total\_weights} \gets \text{total\_weights} \cup \{\tilde{\theta}_k^{(t)}\}$
    \EndFor
    \State $\bar{\theta}^{(t)} \gets \frac{1}{K} \sum_{k=1}^K \tilde{\theta}_k^{(t)}$ 
\EndFor
\State \textbf{Return}: $\bar{\theta}^{(T-1)}$ 
\end{algorithmic}
\end{algorithm}

\subsection{Pruning}
We perform weight pruning on a parameter vector by setting components with absolute values below a threshold to zero (this method can definitely be customized as needed) considering 
$\theta \in \mathbb{R}^n$ as weight vector (dp\_params), where $n$ is the number of parameters and 
$\tau \in \mathbb{R}^+$ as the pruning threshold.

First we create a copy of the weight vector, $\tilde{\theta} = \theta \in \mathbb{R}^n$ (pruned\_weights), and apply pruning as
\[
\tilde{\theta}_i = \begin{cases} 
0 & \text{if } |\theta_i| < \tau \\
\theta_i & \text{otherwise}
\end{cases}, \quad \text{for } i = 1, 2, \dots, n
\]

In a federated learning setup, pruned weights $\tilde{\theta}$ from a client device are sent to a central server, which aggregates weights from multiple clients (e.g., by averaging) to produce an aggregated weight vector $\bar{\theta} \in \mathbb{R}^n$. Two strategies for updating the VQC's initial parameters are considered.
The first strategy assigns the aggregated weights directly to the VQC's initial parameters as
\[
\theta_{\text{initial}} = \bar{\theta}
\]
This approach replaces the current parameters with the server-aggregated weights, ensuring that the VQC reflects the state of the global model after federated aggregation.

The second strategy, employed in a QFL pruning-averaging scheme, updates the VQC parameters by averaging the aggregated weights $\bar{\theta}$ with the current VQC weights $\theta_{\text{current}} \in \mathbb{R}^n$ (which may be pruned or unpruned):
\[
\theta_{\text{initial}} = \frac{\bar{\theta} + \theta_{\text{current}}}{2}
\]

The pruning operation reduces model complexity by setting small weights to zero, which may be particularly beneficial in resource-constrained communications. The direct assignment of aggregated weights ensures alignment with the global model, while the averaging approach balances local and global information, potentially improving robustness in federated learning scenarios.

\subsection{SVD and QKD}
In this approach, we leverage SVD for parameter compression and QKD for secure communication. 
Consider a set of $K$ client devices, each with a VQC parameterized by a weight vector $\theta_k \in \mathbb{R}^n$, where $n$ is the number of parameters and a central server with its own VQC. The goal is to collaboratively optimize a global model by aggregating local weights on $T$ communication rounds, ensuring security through QKD-based encryption and efficiency through SVD compression.

The algorithm operates on $T$ communication rounds $t = \{0, 1, \dots, T-1\}$. 
For each communication round $t$ and each device $k = \{1, 2, \dots, K\}$,
if $t > 0$, initialize the VQC parameters with the aggregated weights from the previous round, $\bar{\theta}^{(t-1)} \in \mathbb{R}^n$ as,
    \[
    \theta_k^{(t)} = \bar{\theta}^{(t-1)}
    \]
Each device $k$ trains its VQC using a local objective function $f_k: \mathbb{R}^n \to \mathbb{R}$, updating the weights $\theta_k^{(t)}$ via a training procedure (e.g. gradient-based optimization). The updated weights are
\[
\theta_k^{(t)} \gets \text{Train}(\theta_k^{(t)}, f_k)
\]

For each device $k$ round $t$,
with the SVD decomposition, we reshape the weight vector $\theta_k^{(t)} \in \mathbb{R}^n$ into a matrix $A_k \in \mathbb{R}^{m \times m}$, where $n = m^2$ (e.g. $m = 4$ for $n = 16$). Compute the SVD as
\[
A_k = U_k \Sigma_k V_k^T
\]
where, $U_k, V_k \in \mathbb{R}^{m \times m}$ are orthogonal matrices, and $\Sigma_k = \text{diag}(\sigma_{k,1}, \sigma_{k,2}, \dots, \sigma_{k,m}) \in \mathbb{R}^{m \times m}$ contains the singular values $\sigma_k = [\sigma_{k,1}, \sigma_{k,2}, \dots, \sigma_{k,m}]^T$.
We convert the singular values $\sigma_k$ into a byte representation, which yields a bit length $n_k$ and a bit string $b_k$. 

With QKD, we generate a random key $s_k \in \{0, 1\}^{n_k}$ and establish shared keys:
\[
(a_k, b_k) = \text{QKD}(b_k)
\]
where, $a_k$ (Sender's key) and $b_k$ (Receiver's key) are derived via QKD operations involving Sender's and Receiver's rotations and measurement results.
Then, for encryption we only encrypt the singular values using Sender's key $a_k$
    \[
    \tilde{\sigma}_k = \text{Encrypt}(\sigma_k, a_k)
    \]
 The tuple $(U_k, \tilde{\sigma}_k, V_k^T)$ represents the compressed and encrypted weights.
Each device $k$ sends its encrypted singular values $\tilde{\sigma}_k$ to the server. The server decrypts using its corresponding key $b_s$ as
\[
\sigma_k' = \text{Decrypt}(\tilde{\sigma}_k, b_s)
\]
The decrypted singular values $\sigma_k'$ are used to reconstruct the weight matrix. For a matrix $A_k \in \mathbb{R}^{m_1 \times m_2}$ (generalizing to non-square cases), select the top $r = \min(\min(m_1, m_2), 2)$ singular values as,
\[
\Sigma_k' = \begin{bmatrix}
\text{diag}(\sigma_{k,1}', \sigma_{k,2}', \dots, \sigma_{k,r}') & 0 \\
0 & 0
\end{bmatrix} \in \mathbb{R}^{m_1 \times m_2}
\]
Reconstruct the weight matrix with
\[
A_k' = U_k \Sigma_k' V_k^T
\]
Then we, flatten to obtain the reconstructed weights as
\[
\theta_k' = \text{Flatten}(A_k') \in \mathbb{R}^n
\]
The reconstructed weights $\theta_k'$ for all devices are collected into a set $\{\theta_k'\}_{k=1}^K$.
The server computes the aggregated weights by averaging the reconstructed weights:
\[
\bar{\theta}^{(t)} = \frac{1}{K} \sum_{k=1}^K \theta_k'
\]

The Algorithm \ref{alg:svd_qkd} integrates SVD for parameter privacy, reducing encryption bottleneck by only encrypting the sigma part, with QKD for secure key exchange, ensuring the confidentiality of model updates.

\begin{algorithm}
\caption{QFL with SVD and QKD Encryption}
\label{alg:svd_qkd}
\begin{algorithmic}[1]
\State \textbf{Input}: $K$ devices with VQC weights $\theta_k \in \mathbb{R}^n$, server VQC, $T$ rounds
\State \textbf{Output}: Aggregated weights $\bar{\theta}^{(T-1)}$
\For{each round $t = \{0, 1, \dots, T-1\}$}
    \State $\bar{\theta}^{(t-1)} \gets \text{null}$ (for $t = 0$)
    \For{each device $k = \{1, 2, \dots, K\}$}
        \State $t_k \gets t$
        \If{$t > 0$}
            \State $\theta_k^{(t)} \gets \bar{\theta}^{(t-1)}$ 
        \EndIf
        \State $\theta_k^{(t)} \gets \text{Train}(\theta_k^{(t)}, f_k)$ 
        \State $A_k \gets \text{Reshape}(\theta_k^{(t)}, (m, m))$ 
        \State $(U_k, \Sigma_k, V_k^T) \gets \text{SVD}(A_k)$, $\Sigma_k = \text{diag}(\sigma_k)$ 
        \State $(n_k, b_k) \gets \text{ConvertToByte}(\sigma_k)$ 
        \State $s_k \gets \text{RandomString}(n_k)$ 
        \State $(a_k, b_k) \gets \text{QKD}(b_k)$ 
        \State $\tilde{\sigma}_k \gets \text{Encrypt}(\sigma_k, a_k)$ 
    \EndFor
    \State Server: $(a_s, b_s) \gets \text{QKD}(b_s)$ 
    \For{each device $k = 1, 2, \dots, K$}
        \State $\sigma_k' \gets \text{Decrypt}(\tilde{\sigma}_k, b_s)$ 
        \State $r \gets \min(\min(m_1, m_2), 2)$ 
        \State $\Sigma_k' \gets \text{Zeros}(m_1, m_2)$, $\Sigma_k'[:r, :r] \gets \text{diag}(\sigma_{k,1}', \dots, \sigma_{k,r}')$
        \State $A_k' \gets U_k \Sigma_k' V_k^T$ 
        \State $\theta_k' \gets \text{Flatten}(A_k')$ 
    \EndFor
    \State $\bar{\theta}^{(t)} \gets \frac{1}{K} \sum_{k=1}^K \theta_k'$ 
\EndFor
\State \textbf{Return}: $\bar{\theta}^{(T-1)}$, 
\end{algorithmic}
\end{algorithm}

\subsection{Condensation}
In this section, we present a theoretical overview of condensation to reduce training data while preserving model performance. 
With the dataset condensation approach, we generate a compact synthetic dataset to reduce computational and communication overhead \cite{yangEfficientDatasetCondensationa, zhaoDATASETCONDENSATIONGRADIENT2021}.

Given a dataset with features $X \in \mathbb{R}^{m \times d}$ and labels $y \in \{0, 1, \dots, C-1\}^m$, where $m$ is the number of samples, $d$ is the feature dimension, and $C$ is the number of classes, condensation produces a synthetic dataset $(X_s, y_s)$ with $|X_s| \ll m$. 
The synthetic features $X_s \in \mathbb{R}^{C \cdot s \times d}$, where $s$ is the number of images per class $c$, are optimized to match the mean embeddings of the real and synthetic data is
\[
\mu_c^{real} = \frac{1}{|I_c|} \sum_{i \in I_c} (x_i W), \quad \mu_{s,c}^{syn} = \frac{1}{|I_{s,c}|} \sum_{i \in I_{s,c}} (x_{s,i} W)
\]
where $I_c$ and $I_{s,c}$ are indices for class $c$ in real and synthetic datasets, respectively, and $W \in \mathbb{R}^{d \times e}$ is a random projection matrix with $e$ as embedding dimension. The objective is to minimize the embedding difference for class $c$ as follows.
\[
\Delta_c = \mu_c^{real} - \mu_{s,c}^{syn} \in \mathbb{R}^d
\]
Then, the loss per class is the squared norm $L_2$:
\[
\mathcal{L}_c = \|\mu_c - \mu_{s,c}\|_2^2
\]
The total loss summed across all classes is:
\[
\mathcal{L}_{total} = \sum_{c=0}^{C-1} \mathcal{L}_c
\]
Now, we compute the gradient with respect to the synthetic images as 
\[
\frac{\partial\mathcal{L}_c}{\partial\vec{x_j}^c} = -\frac{1}{m} \Delta_c W^T
\]
Then, we update the gradient step as, for each synthetic image $\vec{x}_j^c$, update, 
\[
\vec{x}_j^c \leftarrow \vec{x}_j^c - \eta \frac{\partial\mathcal{L}_c}{\partial\vec{x_j}^c}
\]
Additionally, clip to [0,1] as, 
\[
\vec{x}_j^c \leftarrow clip(\vec{x}_j^c, 0, 1).
\]

\begin{algorithm}
\caption{QFL with Dataset Condensation}
\label{alg:dataset-condensation}
\begin{algorithmic}[1]
\State \textbf{Input}: Real images $X \in \mathbb{R}^{n \times h \times w}$, labels $y \in \{0, \dots, C-1\}^n$, images per class $m$, embedding matrix $W \in \mathbb{R}^{hw \times d}$, steps $T$, learning rate $\eta$, batch size $b$
\State \textbf{Output}: Synthetic images $S \in \mathbb{R}^{Cm \times h \times w}$, labels $\tilde{y} \in \{0, \dots, C-1\}^{Cm}$
\State Initialize $S$ with $m$ real images per class, $\tilde{y}$ with corresponding labels
\For{$t = 1$ to $T$}
    \For{each class $c = 0$ to $C-1$}
        \State Sample $b$ real images $X_c$ for class $c$
        \State  Computer Mean Embedding:
        \State For real image, $\mu_r \gets \text{mean}(X_c) \cdot W$ 
        \State For synthetic image, $\mu_s \gets \text{mean}(S_c) \cdot W$ 
        \State Computer $\text{loss} \gets \|\mu_r - \mu_s\|_2^2$
        \State Gradient $g \gets -\frac{2}{m} (\mu_r - \mu_s) \cdot W^T$
        \State $S_c \gets S_c - \eta \cdot g$
        \State Clip $S_c \gets \text{clip}(S_c, 0, 1)$ 
    \EndFor
\EndFor
\State \textbf{Return}: $S$, $\tilde{y}$
\end{algorithmic}
\end{algorithm}

\section{Experimentation}

\subsection{Setup}

\textbf{Dataset.}
In this study, the datasets used include the IRIS dataset, the MNIST dataset, and a genomic dataset.
The IRIS dataset consists of 3 classes, each representing a different iris species (Setosa, Versicolour, Virginica), with 50 samples per class.  
The MNIST dataset contains images of handwritten digits, with 60,000 samples in the training set and 10,000 samples in the test set, and includes 10 distinct class labels.  
The genomic dataset comprises genomic data annotated with 2 class labels. 
The genomic data\footnote{\url{https://github.com/ML-Bioinfo-CEITEC/genomic_benchmarks}} is a dataset designed for classifying genomic sequences.  
Although the repository includes multiple datasets, for our experiments we use the ``demoHumanOrWorm" subset.  
Since the data consists of genome sequences such as ``ATGC.", it is preprocessed and transformed into a numerical representation.  
To keep the experimental analysis tractable, we apply PCA with 4 components to perform feature dimensionality reduction.

\textbf{Tools.}
We use various tools to perform our experimental analysis, including Qiskit, IBM PCA\_DP library, 
SVD library 
etc. with integration to various protocols and approaches proposed in this work.
In terms of quantum classifier, we utilized the VQC with COBYLA optimizer, ZFeatureMap as a feature map of reps=1, RealAmplitudes as an ansatz of reps=3 and AerSimulator as backend.

\textbf{Metrics.}
In this work, we evaluate the following metrics.
\begin{enumerate}
    \item Global Model Adaptability (G+): After all local devices train their local model, once we generate a global average model, we use that model to fit into unseen validation set of server device and test on unseen test set of the server as well. The model fitted or adapted on server side is still not used to update the local devices and is only used to study the adaptability or learning capability of the global model on unseen dataset.
    \item Global Model Prediction Results (Prediction Model - Pred.): This result is the prediction capability of the freshly averaged global model and is different from the previous adapted or fine tuned model.
    This (freshly) averaged model is the one used by local devices to update their local model.
    \item Local Model Performance (Local Model): Once the local devices receive an updated global model, they update their model and train the model. The freshly trained model is then again tested on the local devices.
    \item Communication time: This refers to the total time required to complete a single communication round which involves, training, averaging, and updating of local models again.
\end{enumerate}

\subsection{Results}

\begin{figure}[!htbp]
    \centering
    \begin{subfigure}[b]{0.3\columnwidth}
        \centering
       \includegraphics[width=\columnwidth]{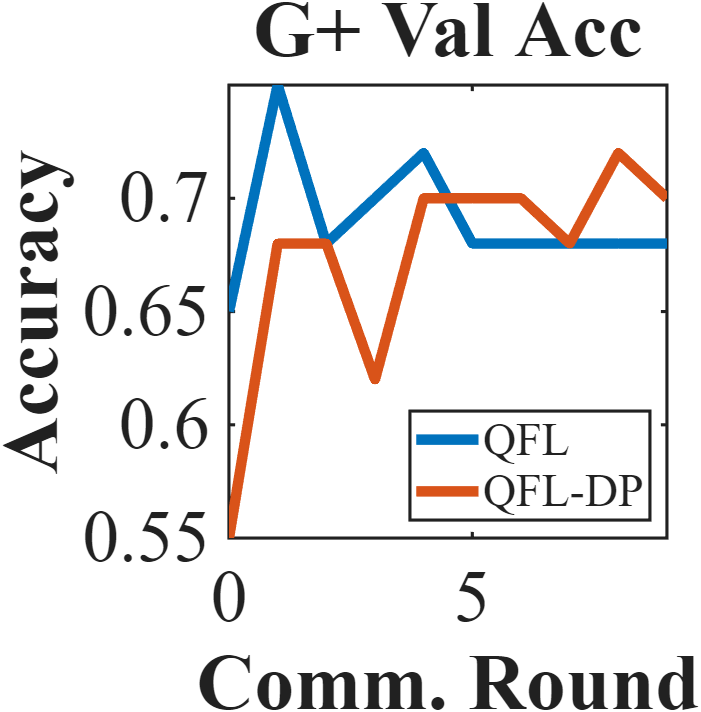}
    \caption{G+ Test}
    \label{fig:server_val_acc_dp_pca}
    \end{subfigure}
    \begin{subfigure}[b]{0.3\columnwidth}
        \centering
       \includegraphics[width=\columnwidth]{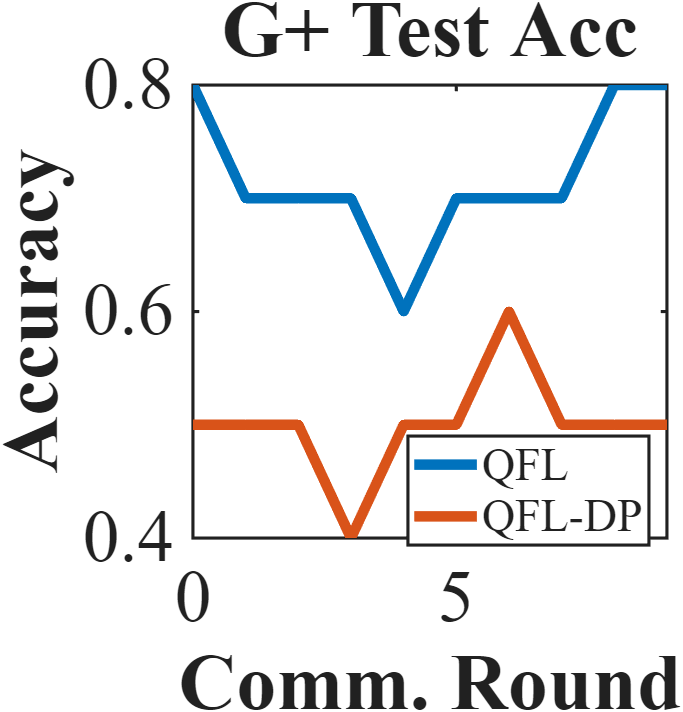}
    \caption{G+ Val}
    \label{fig:server_test_acc_dp_pca}
    \end{subfigure}
  \begin{subfigure}[b]{0.32\columnwidth}
        \centering
      \includegraphics[width=\columnwidth]{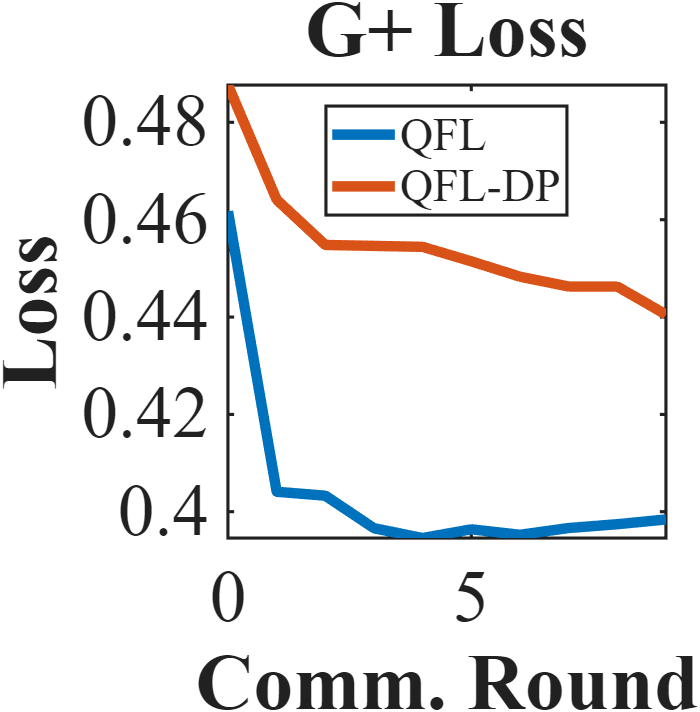}
    \caption{G+ Loss}
    \label{fig:server_val_loss_dp_pca}
    \end{subfigure}
    \begin{subfigure}[b]{0.3\columnwidth}
        \centering
       \includegraphics[width=\columnwidth]{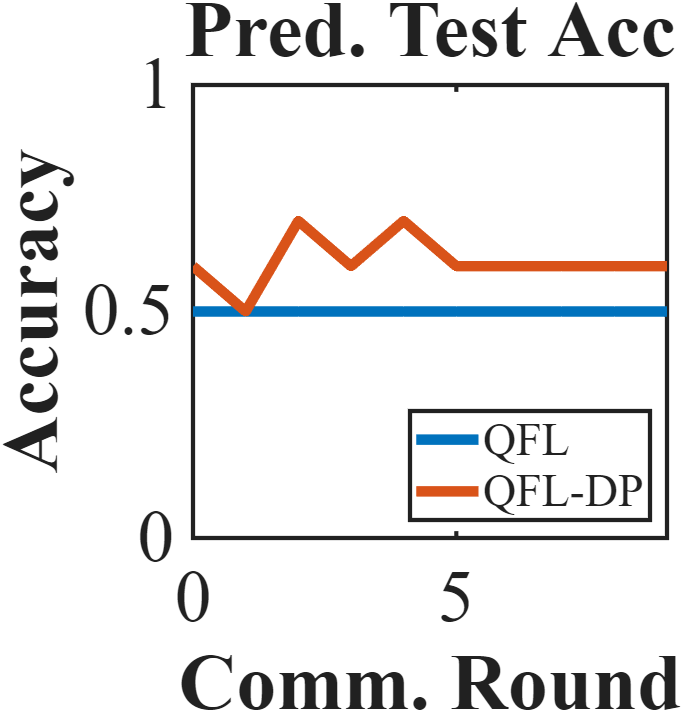}
    \caption{Pred. Test}
    \label{fig:prediction_test_dp_pca}
    \end{subfigure}
    \begin{subfigure}[b]{0.3\columnwidth}
        \centering
       \includegraphics[width=\columnwidth]{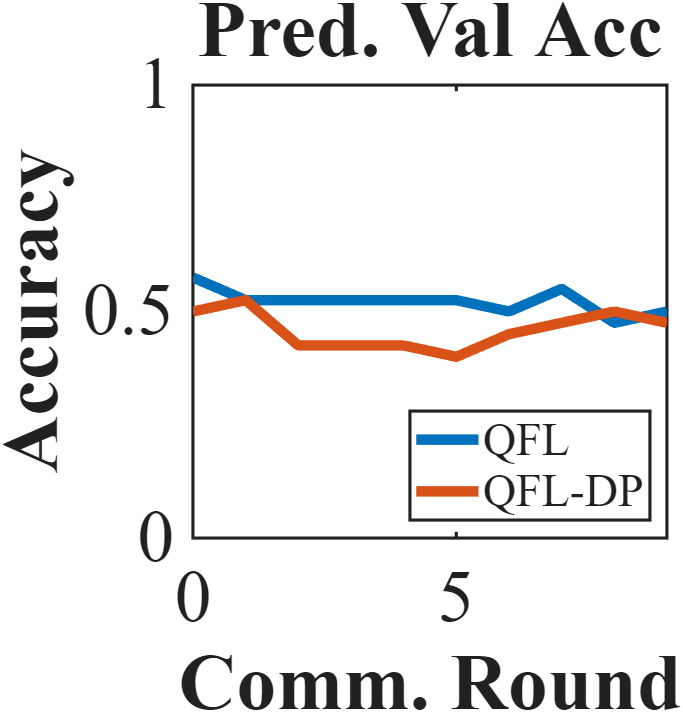}
    \caption{Pred. Val}
    \label{fig:prediction_val_dp_pca}
    \end{subfigure}
  \begin{subfigure}[b]{0.3\columnwidth}
        \centering
      \includegraphics[width=\columnwidth]{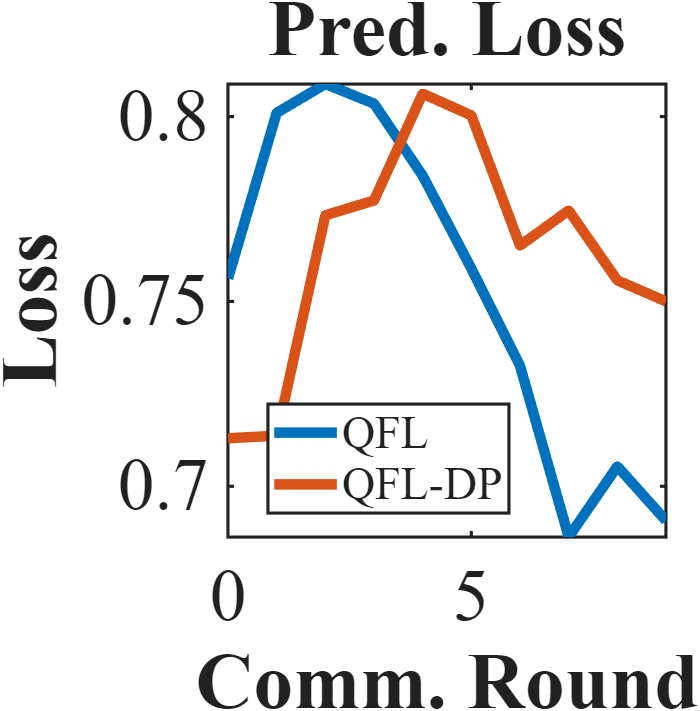}
    \caption{Pred. Loss}
    \label{fig:prediction_loss_dp_pca}
    \end{subfigure}
     \begin{subfigure}[b]{0.3\columnwidth}
        \centering
       \includegraphics[width=\columnwidth]{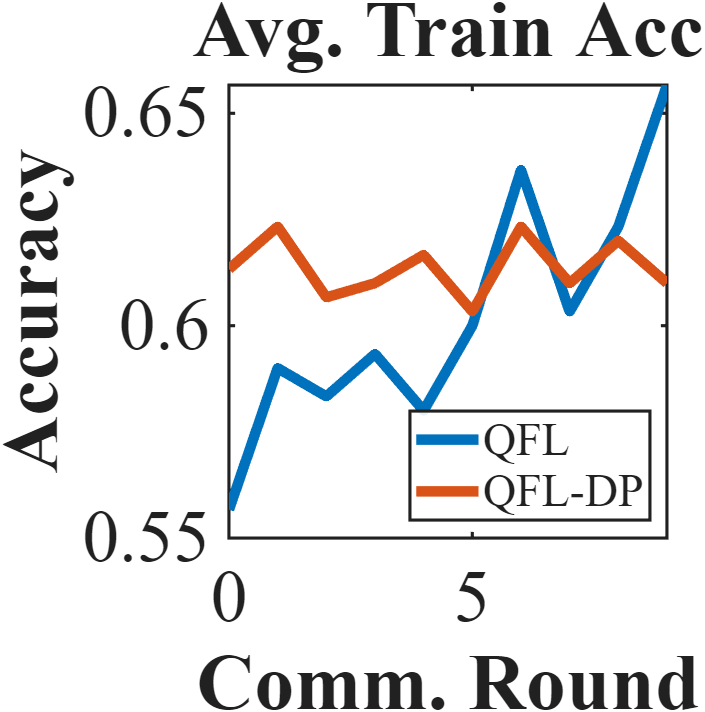}
    \caption{Avg. Train Acc}
    \label{fig:devices_train_acc_dp_pca}
    \end{subfigure}
    \begin{subfigure}[b]{0.3\columnwidth}
        \centering
       \includegraphics[width=\columnwidth]{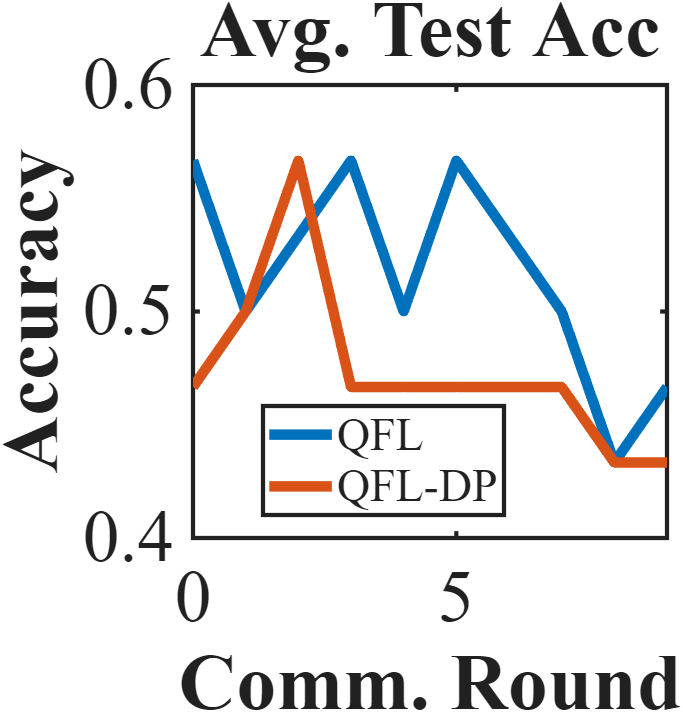}
    \caption{Avg. Test Acc}
    \label{fig:devices_test_acc_dp_pca}
    \end{subfigure}
    \centering
    \begin{subfigure}[b]{0.3\columnwidth}
        \centering
       \includegraphics[width=\linewidth]{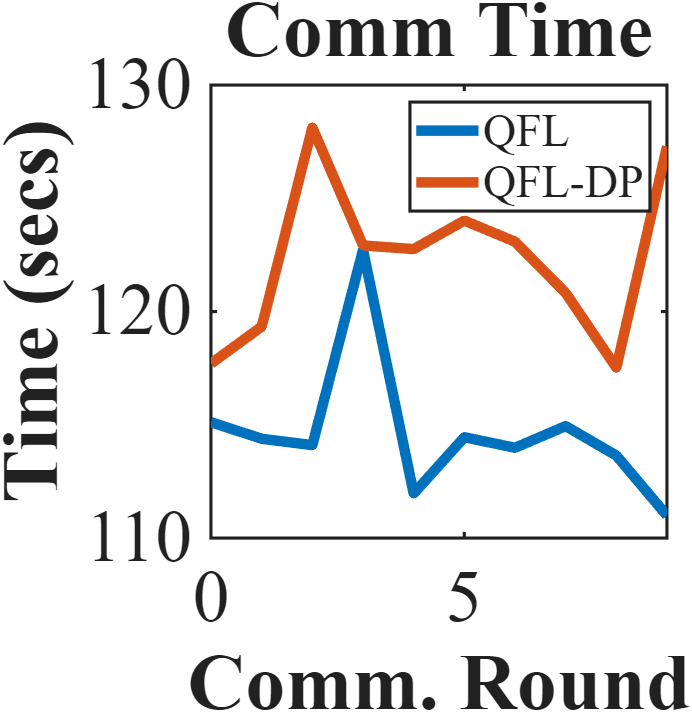}
    \caption{Comm Time}
    \label{fig:comm_time_dp_pca}
    \end{subfigure}   
    \caption{Global Model Adaptation (G+), Prediction (Pred.), Average Devices Performance (Avg.): PCA vs PCA\_DP}
    \label{fig:performance_pc_dp_pca}
\end{figure}

\subsubsection{PCA\_DP}
In this experiment, we used a genomic dataset of 2000 samples for devices and 200 samples for server.
We limit the number of devices to 3 because we want to observe the impact on accuracy due to the integration of PCA.\_DP instead of the standard PCA.
The number of components for PCA is 4, the optimizer max iteration is 100, for PCA\_DP we use epsilon=1.0, bounds = (0.0, 1.0), data norm = 1.0.
The results are presented in Figure \ref{fig:performance_pc_dp_pca}.
We can clearly observe the impact of implemented DP with PCA in almost all results showing that performance is deteriorated in all results like for the performance of the adapted model (G+ Val Acc, Test Acc and Loss) in Figures \ref{fig:server_val_acc_dp_pca}, \ref{fig:server_test_acc_dp_pca} and \ref{fig:server_val_loss_dp_pca}.
However, the prediction results in terms of test accuracy are better with DP PCA (QFL-DP) than with QFL and similar in terms of validation prediction result as in  Figures \ref{fig:prediction_test_dp_pca} and \ref{fig:prediction_val_dp_pca}.
This contradicts to the notion that adding noise (DP) with PCA isn't impacting the performance.
With average device performance as in Figures \ref{fig:devices_train_acc_dp_pca} and \ref{fig:devices_test_acc_dp_pca}, the default QFL is better with QFL-DP having more communication bottleneck as seen in Figure \ref{fig:comm_time_dp_pca}.

\subsubsection{DP Noise Variations}
In this experiment, we compare various methods with no noise (QFL), with noise (QFL-N) with variations (QFL-N-G or QFL-N-L where G is for Gaussian noise and L is for Laplace noise), with PCA-DP (QFL-DP), with noise and PCA-DP (QFL-DP-N) with variations (QFL-DP-N-L or QFL-DP-N-G).
The results are presented in Table \ref{tab:noise_dp_variations}.
The experiments were performed with both both IRIS and the genomic data set.
DP PCA is applied on the dataset whereas noise is added on the model parameters of Laplace or Gaussian Type with epsilon value = 1.0.
In terms of adapted model (G+) results, we find not much difference between QFL and other variations with noise, DP etc. This is due to the G+ model already adapted.
In general, there is no massive impact due to noise or DP methods.
With Genomic dataset with only 3 devices (Genomic), G+ Test accuracy is better with DP.
Even with top performing devices, communication time, the results are somehow similar with varying methods even with Genomic dataset experiment (Genomic 10) with 10 devices.
However, with a straight comparison between QFL and QFL-DP, we see better results with QFL, however, the best results are equal from both methods.
While with DP and noise, we have extended our privacy guaranties, with these results we can conclude the impact of noise is minimal and usable.

\begin{table*}[!htbp]
\centering
\caption{Performance between QFL with Noise and DP and their combinations on IRIS and Genomic Dataset.}
\label{tab:noise_dp_variations}
\resizebox{0.9\textwidth}{!}{
\small
\begin{tabular}{
    l p{2cm}
    *{3}{p{0.55cm}}   
    *{3}{p{0.55cm}}   
    *{3}{p{0.55cm}}   
    *{3}{p{0.55cm}}   
    *{2}{p{2cm}}    
    p{0.7cm}         
}
\toprule
\multirow{2}{*}{Dataset} & \multirow{2}{*}{Model}
& \multicolumn{3}{c}{G+ Val Acc}
& \multicolumn{3}{c}{G+ Test Acc}
& \multicolumn{3}{c}{Avg. Devices Train}
& \multicolumn{3}{c}{Avg. Devices Test}
& \multicolumn{2}{c}{Top Device Accuracy}
& \multirow{2}{*}{Comm} \\
\cmidrule(lr){3-5} \cmidrule(lr){6-8} \cmidrule(lr){9-11} \cmidrule(lr){12-14} \cmidrule(lr){15-16}
& & Avg & Final & Max
& Avg & Final & Max
& Avg & Final & Max
& Avg & Final & Max
& Test Acc& Train Acc& Time\\
\toprule
\multirow{6}{*}{Iris}
 & QFL        & 0.548 & 0.58 & 0.58 & 0.67 & 0.67 & 0.67 & 0.46 & 0.53 & 0.53 & 0.51 & 0.67 & 0.67 & R9-D0 (0.89) & R8-D2 (0.67) & 116.79 \\
 & QFL-N-G    & 0.564 & 0.58 & 0.58 & 0.67 & 0.67 & 0.67 & 0.49         & 0.52         & 0.52         & 0.55         & 0.59 & 0.59 & R0-D1 (0.89) & R2-D2 (0.61) & 116.77 \\
 & QFL-N-L    & 0.524 & 0.58 & 0.58 & 0.67 & 0.67 & 0.67 & 0.46 & 0.52         & 0.52         & 0.47         & 0.52 & 0.55 & R0-D1 (0.78) & R9-D2 (0.61) & 118.76 \\
 & QFL-DP     & 0.58  & 0.58 & 0.58 & 0.67 & 0.67 & 0.67 & 0.49         & 0.45         & 0.53         & 0.50         & 0.44 & 0.59 & R0-D1 (0.67) & R1-D2 (0.64) & 118.27 \\
 & QFL-DP-N-L & 0.572 & 0.58 & 0.58 & 0.67 & 0.67 & 0.67 & 0.45         & 0.52         & 0.52         & 0.52         & 0.59 & 0.63 & R0-D1 (0.89) & R7-D2 (0.64) & 119.99 \\
 & QFL-DP-N-G & 0.58  & 0.58 & 0.58 & 0.67 & 0.67 & 0.67 & 0.43         & 0.43         & 0.49         & 0.49         & 0.56 & 0.59 & R3-D1 (0.89) & R3-D1 (0.56) & 118.18 \\
\midrule
\multirow{6}{*}{Genomic}
 & QFL        & 0.697 & 0.72 & 0.75 & 0.41 & 0.4  & 0.5  & 0.71 & 0.76 & 0.76 & 0.38 & 0.27 & 0.5  & R2-D0 (0.60) & R1-D0 (0.80) & 114.18\\
 & QFL-N-G    & 0.731 & 0.75 & 0.8  & 0.4  & 0.4  & 0.5  & 0.71 & 0.73 & 0.73 & 0.31 & 0.37 & 0.4  & R0-D0 (0.50) & R5-D1 (0.85) & 118.3 \\
 & QFL-N-L    & 0.751 & 0.8 & 0.8 & 0.48 & 0.5 & 0.6 & 0.71 & 0.7  & 0.76 & 0.45 & 0.3  & 0.53 & R1-D2 (0.70) & R2-D1 (0.80) & 117.31 \\
 & QFL-DP     & 0.715 & 0.7  & 0.78 & 0.67 & 0.7 & 0.8 & 0.74 & 0.76 & 0.76 & 0.42 & 0.4  & 0.43 & R1-D1 (0.60) & R4-D0 (0.82) & 116.66 \\
 & QFL-DP-N-L & 0.658 & 0.68 & 0.75 & 0.61 & 0.6  & 0.8 & 0.73 & 0.75 & 0.77 & 0.5  & 0.53 & 0.57 & R4-D0 (0.70) & R5-D1 (0.88) & 115.88 \\
 & QFL-DP-N-G & 0.698 & 0.78 & 0.78 & 0.44 & 0.4  & 0.6  & 0.74 & 0.73 & 0.77 & 0.42 & 0.43 & 0.47 & R0-D0 (0.50) & R3-D0 (0.88) & 116.26 \\
\midrule
\multirow{6}{*}{Genomic 10}
 & QFL & 0.636 & 0.66 & 0.68 & 0.488 & 0.53 & 0.55 & 0.58 & 0.59 & 0.60 & 0.52 & 0.52 & 0.57 & R0-D4 (0.72) & R9-D9 (0.68) & 798.36 \\
 & QFL-N-G & 0.604 & 0.58 & 0.62 & 0.572 & 0.6 & 0.65 & 0.6 & 0.61 & 0.61 & 0.5 & 0.51 & 0.54 & R6-D3 (0.72) & R1-D4 (0.67) & 801.29 \\
 & QFL-N-L & 0.587 & 0.55 & 0.62 & 0.479 & 0.45 & 0.55 & 0.6 & 0.6 & 0.61 & 0.52 & 0.53 & 0.55 & R4-D0 (0.68) & R4-D7 (0.67) & 802.29 \\
 & QFL-DP & 0.597 & 0.61 & 0.62 & 0.617 & 0.57 & 0.7 & 0.61 & 0.62 & 0.62 & 0.52 & 0.48 & 0.56 & R1-D7 (0.72) & R4-D3 (0.70) & 784.84 \\
 & QFL-DP-N-L & 0.601 & 0.6 & 0.62 & 0.6 & 0.6 & 0.65 & 0.58 & 0.59 & 0.59 & 0.53 & 0.53 & 0.55 & R4-D1 (0.68) & R2-D1 (0.65) & 783.88 \\
 & QFL-DP-N-G & 0.594 & 0.65 & 0.65 & 0.56 & 0.55 & 0.65 & 0.59 & 0.59 & 0.61 & 0.51 & 0.49 & 0.52 & R5-D8 (0.72) & R7-D0 (0.65) & 772.53 \\
 \midrule
\multirow{2}{*}{Iris}
 & QFL        & \textbf{0.58} & \textbf{0.58} & \textbf{0.58} & \textbf{0.67} & \textbf{0.67} & \textbf{0.67} & \textbf{0.63} & \textbf{0.64} & \textbf{0.64} & \textbf{0.74} & \textbf{0.74} & \textbf{0.74} & R0-D2 (0.89) & R1-D0 (0.69) & \textbf{24.41} \\
 & QFL-DP     & 0.492 & 0.50 & 0.50 & 0.297 & 0.33 & 0.33 & 0.56 & 0.56 & 0.58 & 0.56 & 0.56 & 0.59 & R0-D2 (0.89) & R1-D0 (0.69) & 24.48 \\
 \midrule
\multirow{2}{*}{Genomic}
 & QFL & \textbf{0.69} & \textbf{0.68} & \textbf{0.75} & \textbf{0.72} & \textbf{0.8} & \textbf{0.8} & \textbf{0.6} & \textbf{0.66} & \textbf{0.66} & \textbf{0.52} & \textbf{0.47} & \textbf{0.57} & R3-D0 (0.80) & R5-D1 (0.72) & \textbf{114.63} \\
 & QFL-DP & 0.673 & 0.7 & 0.72 & 0.5 & 0.5 & 0.6 & 0.61 & 0.61 & 0.62 & 0.47 & 0.43 & 0.57 & R0-D0 (0.70) & R2-D0 (0.72) & 122.36 \\
\bottomrule
\multicolumn{17}{p{\linewidth}}{\vspace{0.15ex}\par\small
 Top devices (R - D) - R communication round number, D Device Number; Max - Highest Accuracy, Final - Accuracy at the end of all communication rounds, Avg - Average results in each communication round for all devices, Avg-Avg - Average of all average accuracies across all communication rounds for devices.
} \\
\end{tabular}
}
\end{table*}

\subsubsection{AQGD DP}
\begin{figure}[!htbp]
    \centering
    \begin{subfigure}[b]{0.24\columnwidth}
        \centering
       \includegraphics[width=\columnwidth]{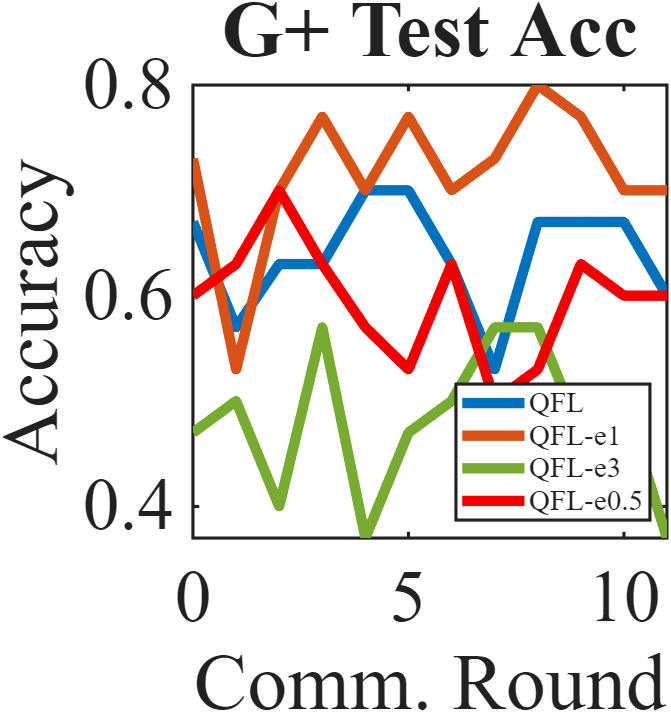}
    \caption{G+ Test}
    \label{fig:server_test_aqgd}
    \end{subfigure}
    \begin{subfigure}[b]{0.24\columnwidth}
        \centering
       \includegraphics[width=\columnwidth]{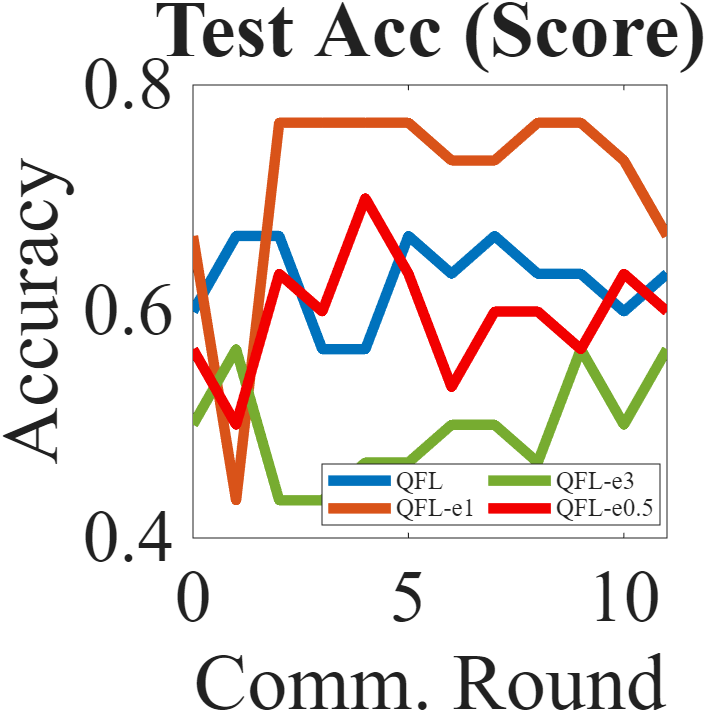}
    \caption{Score}
    \label{fig:server_test_score_aqgd}
    \end{subfigure}
       \begin{subfigure}[b]{0.24\columnwidth}
        \centering
       \includegraphics[width=\columnwidth]{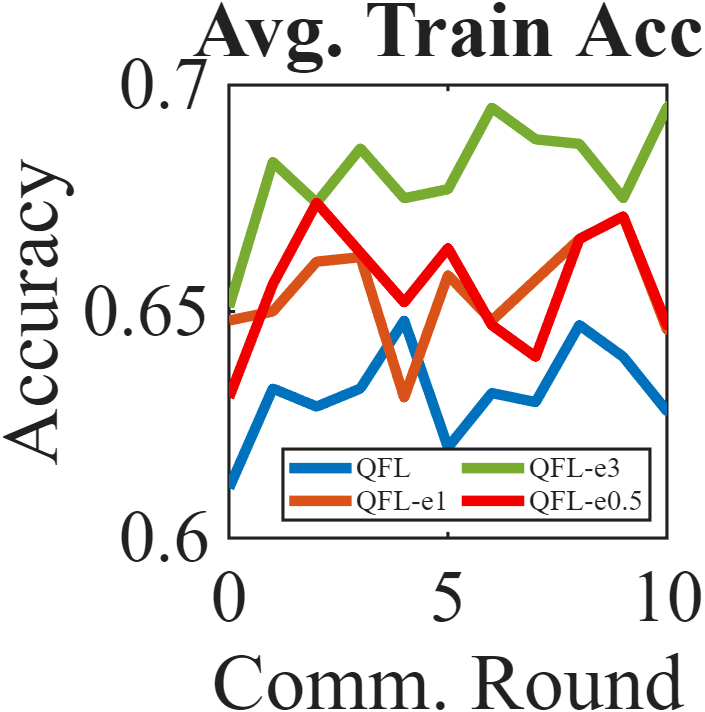}
    \caption{Avg. Train}
    \label{fig:train_acc_aqgd}
    \end{subfigure}
    \begin{subfigure}[b]{0.24\columnwidth}
        \centering
       \includegraphics[width=\columnwidth]{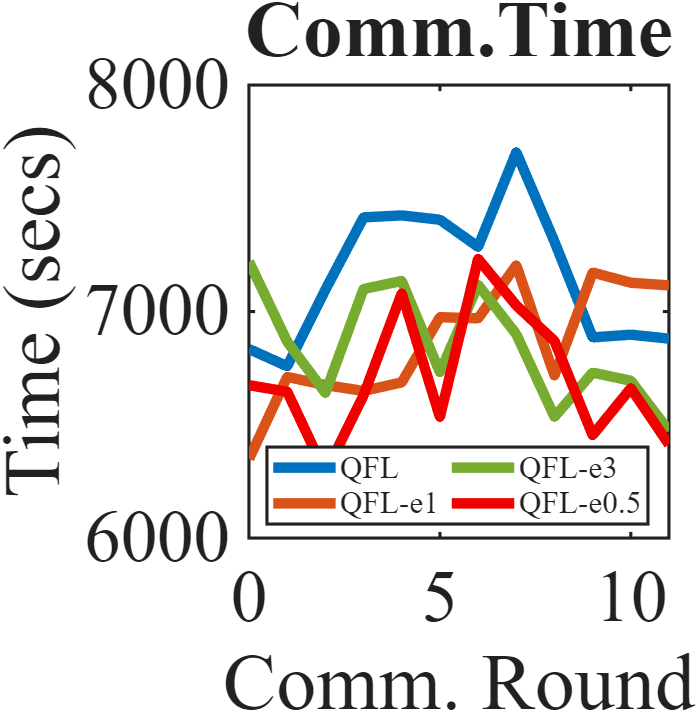}
    \caption{Comm Time}
    \label{fig:comm_time_aqgd}
    \end{subfigure}
    \caption{AQGD vs AQGP\_DP; QFL-AQGD\_DP variations with QFL-AQDP}
    \label{fig:performance_aqgd_vs_default}
\end{figure}

In this experiment, we customize the default AQGD in Qiskit to add DP to its gradient descent optimization.
Dataset used is Genomic, 1000 samples for devices, 150 for server,  10 devices, data distributed among devices.
For privacy parameters for AQGD, we select various variations for epsilon = 1, 3.0, 0.5.
Delta and sensitivity values are set to 1e-5 and 1.0 respectively.
AQGD max iteration is set to 100, run on AerSimulator.
We present the results in Figure \ref{fig:performance_aqgd_vs_default} and Table \ref{tab:performance_aqgd_vs_default}.

The method QFL-e1 achieves the highest performance in terms of G+ test accuracy (Figure \ref{fig:server_test_aqgd}) as well as Server Score (Figure \ref{fig:server_test_score_aqgd}). 
The Server Score is computed using the scoring functionality of the Qiskit VQC model, applied to a model initialized with the newly averaged global parameters on the server test set, prior to any adaptation, in other words, without using the G+ model.
Thus, the result of the G+ model and the result of the score are different.
However, at the device level, we can see that the results are better with QFL-e3 and QFL-e0.5 as in Table \ref{tab:performance_aqgd_vs_default}.
Thus, given its ability to preserve gradient privacy while keeping global results and local model performance comparable, AQGD DP appears to be advantageous.. 
Overall communication time is roughly the same, with minor variations that are likely due to factors such 
as the Google Colab environment or internet configuration, rather than being directly caused by the DP methods themselves.

\begin{table}[htbp]
\centering
\caption{QFL Performance Summary (Genomics): QFL vs QFL-AQGD}
\label{tab:performance_aqgd_vs_default}
\resizebox{\columnwidth}{!}{
\begin{tabular}{|l|cc|cc|cc|cc|cc|c|}
\toprule
\textbf{Model} &
\multicolumn{4}{c|}{\textbf{G+ Accuracy}} &
\multicolumn{4}{c|}{\textbf{Local Device}} &
\multicolumn{2}{c|}{\textbf{Score}} &
\textbf{Comm (s)} \\
\cmidrule(lr){2-5} \cmidrule(lr){6-9} \cmidrule(lr){10-11} \cmidrule(lr){12-12}
&
\multicolumn{2}{c|}{Val} &
\multicolumn{2}{c|}{Test} &
\multicolumn{2}{c|}{Train Acc} &
\multicolumn{2}{c|}{Test Acc} &
Avg & Final & \\
\cmidrule(lr){2-3} \cmidrule(lr){4-5} \cmidrule(lr){6-7} \cmidrule(lr){8-9}
& Avg    & Final  & Avg    & Final  & Avg    & Final  & Avg    & Final  &     &     & \\
\midrule
QFL      & 0.6917 & 0.6800 & 0.6392 & 0.6000 & 0.6328 & 0.6430 & 0.5508 & 0.5600 & 0.6278 & 0.6333 & 7157.92 \\
QFL-e1   & 0.5950 & 0.6200 & 0.7167 & 0.7000 & 0.6536 & 0.6400 & 0.5935 & 0.5250 & 0.7139 & 0.6667 & 6862.62 \\
QFL-e3   & 0.6508 & 0.6500 & 0.4800 & 0.3700 & 0.6808 & 0.6600 & 0.6250 & 0.8000 & 0.4972 & 0.5667 & 6845.34 \\
QFL-e0.5 & 0.6475 & 0.6600 & 0.5958 & 0.6000 & 0.6562 & 0.6575 & 0.6071 & 0.7125 & 0.5972 & 0.6000 & 6709.17 \\
\bottomrule
\end{tabular}
}
\end{table}

\subsubsection{Model Privacy}
In this set of experiments, we compare between QFL, QFL\_pruning, QFL\_pruning\_avg, QFL\_SVD\_QKD\_Encryption, QFL\_QKD methods.
In QFL\_pruning, we just prune a part of the the model parameters to a a certain value such as zero (i.e. a column values of a weight matrix vector).
In terms of updating the local model after the server sends the global model to the clients, 
there are two approaches, with one, the global model is directly assigned to the local client model (prun) while with another, we average the global model 
and the previous local model and update the local model for the client (prun\_avg).
Also, we compare another approach where we perform SVD on model parameters, only encrypt the sigma part with QKD key so that we don't need bigger key (svd\_qkd).
Similarly, we have also included results with only QKD encryption for privacy (qkd).
The datasets used are
Genomics with 2000-150 train and test set, 100 max iter and 10 devices.
With IRIS 150 samples with 3 Devices.

\begin{figure}[!htbp]
    \centering
    \begin{subfigure}[b]{0.24\columnwidth}
        \centering
       \includegraphics[width=\columnwidth]{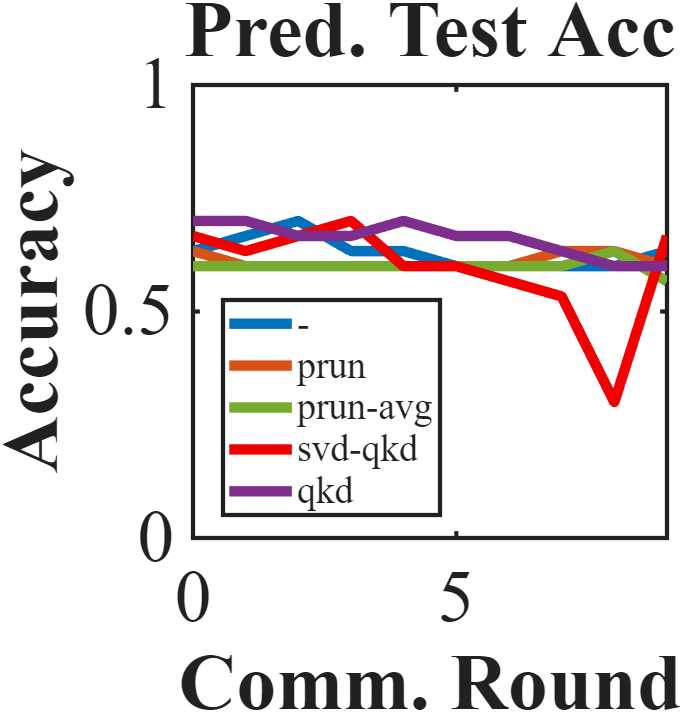}
    \caption{Test Gen}
    \label{fig:prediction_test_acc_prun_genomic}
    \end{subfigure}
    \begin{subfigure}[b]{0.24\columnwidth}
        \centering
       \includegraphics[width=\columnwidth]{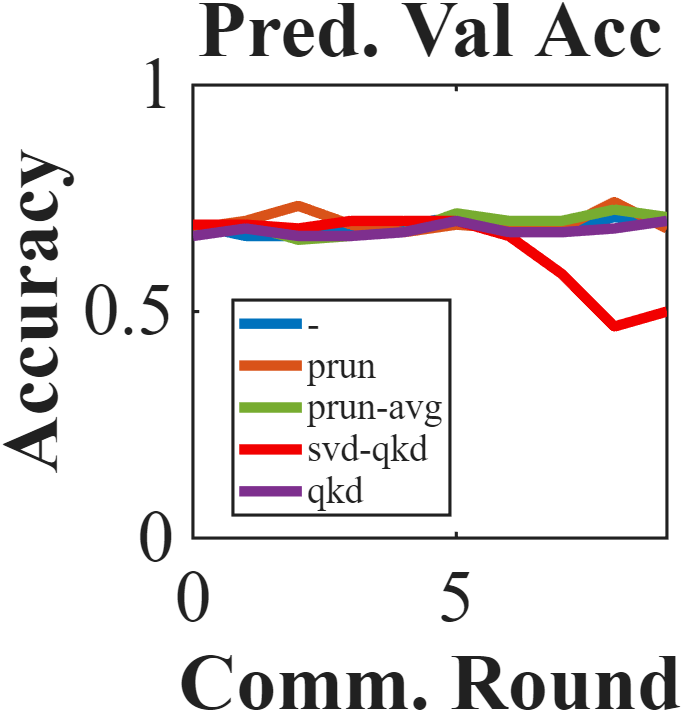}
    \caption{Val Gen}
    \label{fig:prediction_val_acc_prun_genomic}
    \end{subfigure}
    \begin{subfigure}[b]{0.24\columnwidth}
        \centering
       \includegraphics[width=\columnwidth]{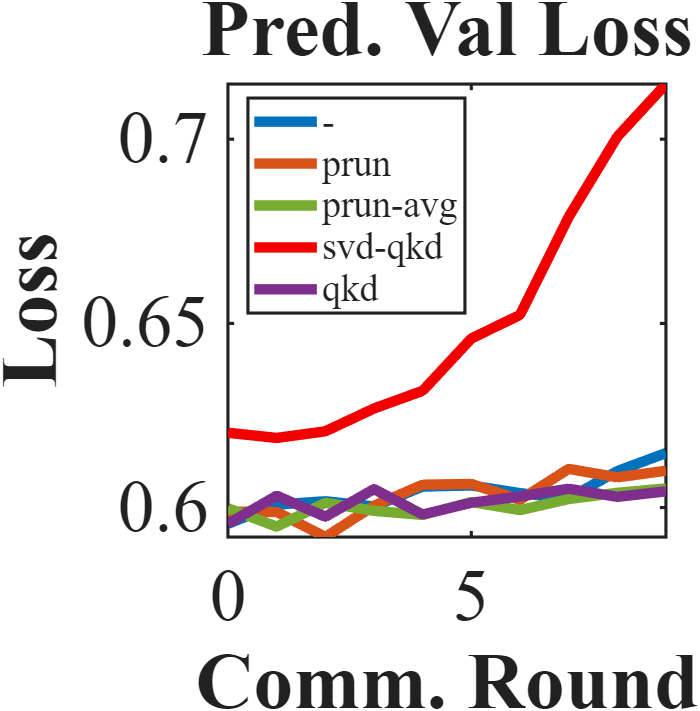}
    \caption{Loss Gen}
    \label{fig:prediction_val_loss_prun_genomic}
    \end{subfigure}
       \begin{subfigure}[b]{0.24\columnwidth}
        \centering
       \includegraphics[width=\columnwidth]{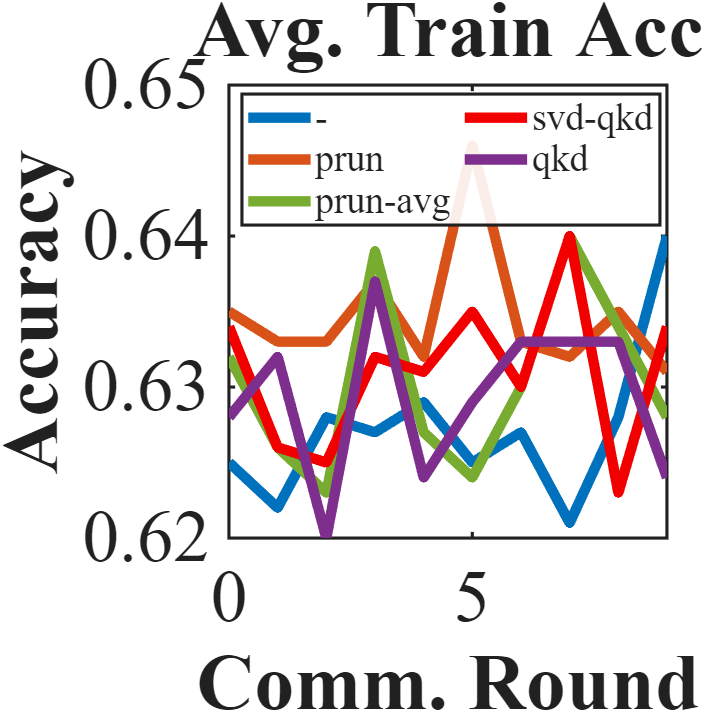}
    \caption{Avg. Train}
    \label{fig:average_devices_train_acc_prun_genomic}
    \end{subfigure}
     \begin{subfigure}[b]{0.24\columnwidth}
        \centering
       \includegraphics[width=\columnwidth]{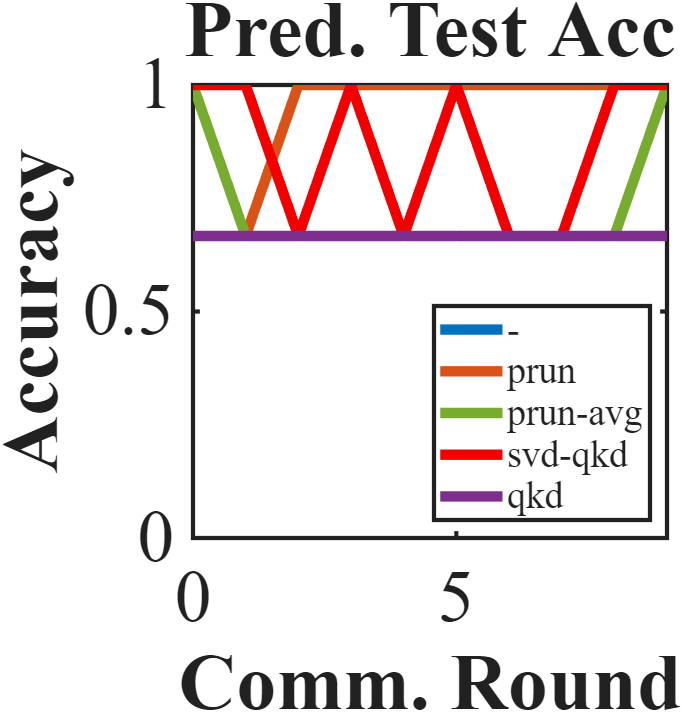}
    \caption{Test - IRIS}
    \label{fig:prediction_test_acc_prun_iris}
    \end{subfigure}
     \begin{subfigure}[b]{0.24\columnwidth}
        \centering
       \includegraphics[width=\columnwidth]{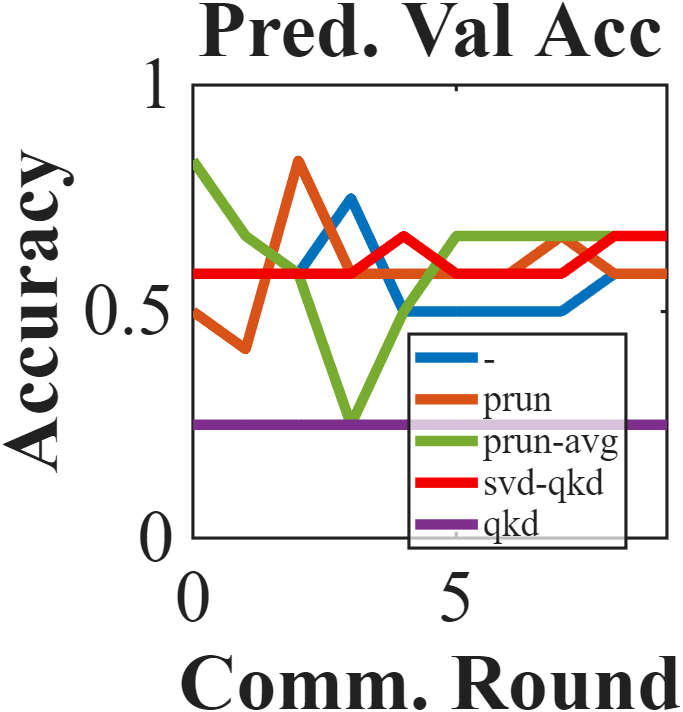}
    \caption{Val - IRIS}
    \label{fig:prediction_val_acc_prun_iris}
    \end{subfigure}
     \begin{subfigure}[b]{0.24\columnwidth}
        \centering
       \includegraphics[width=\columnwidth]{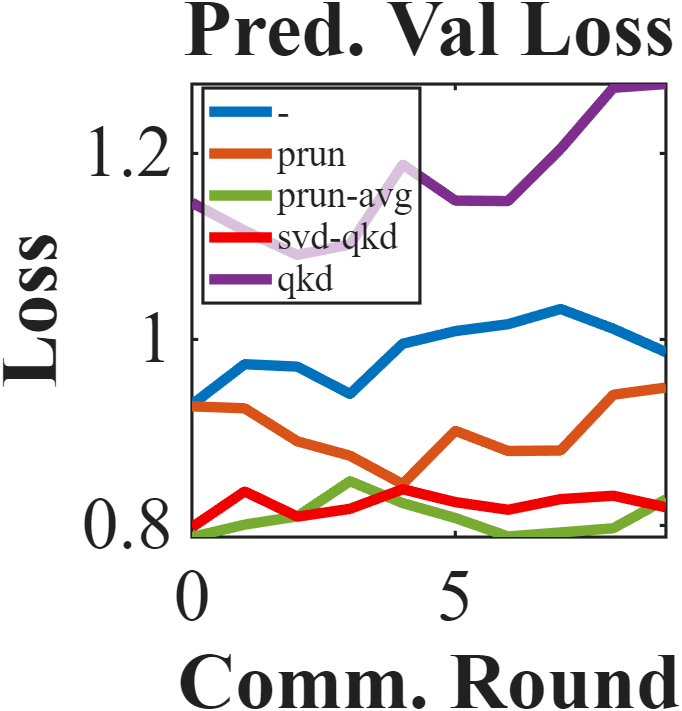}
    \caption{Loss - IRIS}
    \label{fig:prediction_val_loss_prun_iris}
    \end{subfigure}
       \begin{subfigure}[b]{0.24\columnwidth}
        \centering
       \includegraphics[width=\columnwidth]{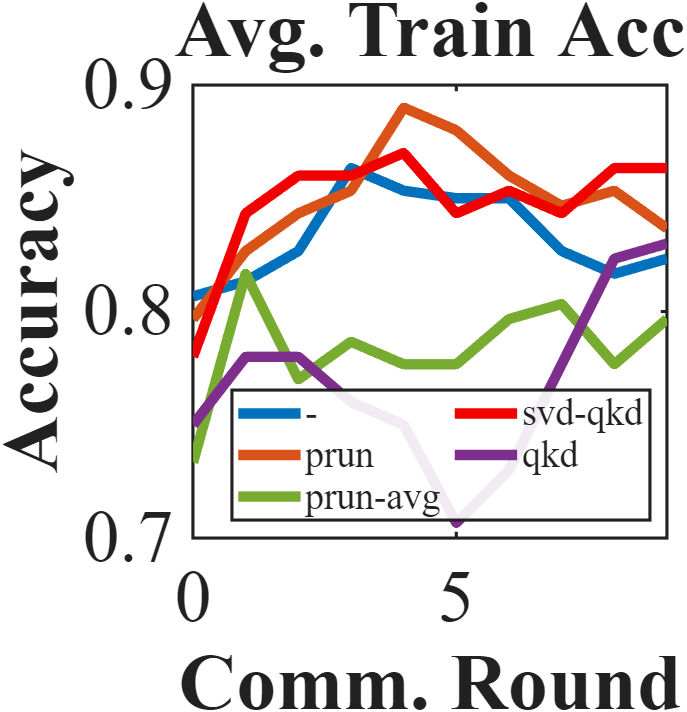}
    \caption{Avg. Train}
    \label{fig:average_devices_train_acc_prun_iris}
    \end{subfigure}
    \caption{Prediction Accuracy and Devices Performance: IRIS/Genomic Dataset; 100 maxiter; Various model privacy methods}
    \label{fig:model_privacy_pruning}
\end{figure}

In Figure \ref{fig:model_privacy_pruning}, we can observe prediction results, and average devices performance for Genomic and IRIS dataset.
The method svd-qkd suffers most in terms of Genomic data as in Figures \ref{fig:prediction_test_acc_prun_genomic} and \ref{fig:prediction_val_acc_prun_genomic}.
Whereas with IRIS dataset, qkd performs the worst.
These highlights there could be impact of QKD overall in the process of verification, model encryption etc.
Since, QKD is not intended to impact on the performance as it just encrypts and decrypts the model parameters without any corruption to the model parameters.
Best method for average devices is with only pruning (prun) for Genomic (Figure \ref{fig:average_devices_train_acc_prun_genomic}) and for IRIS it is skd-qkd (Figure \ref{fig:average_devices_train_acc_prun_iris}).
Whereas, QFL is not the best performing method in this experiment.
This shows that there is no impact on the performance of the system with the integration of protocols like QKD, SVD, pruning, etc.
The results are also presented in Table \ref{tab:pruning}.
The main implication of the results in the table is that the results are somehow comparable between QFL and QFL with additional privacy protocols.
This is good as we are getting privacy guarantee without much performance bottlenecks.

\begin{table}[htbp]
\centering
\caption{Results on IRIS/Genomic Dataset due to pruning, skd, QKD etc.}
\label{tab:pruning}
\small
\resizebox{\columnwidth}{!}{
\begin{tabular}{ll|ccc|c|cc|cc}
\toprule
\multirow{2}{*}{\textbf{Dataset}} &
\multirow{2}{*}{\textbf{Model}} &
\multicolumn{3}{c|}{\textbf{Prediction Results}} &
\multicolumn{1}{c|}{\textbf{Comm Time}} &
\multicolumn{2}{c|}{\textbf{Device Performance}} &
\multicolumn{2}{c}{\textbf{G+ Performance}} \\
\cmidrule(lr){3-5} \cmidrule(lr){6-6} \cmidrule(lr){7-8} \cmidrule(lr){9-10}
& & Avg & Final & Top & Time (s) & Avg Train & Final Train & Val Acc & Test Acc \\
\midrule
\multirow{5}{*}{IRIS}
& -         & 0.7000 & 0.6667 & 1.0000 & 107.66 & 0.83 & 0.82 & 0.7820 & 1.0000 \\
& prun      & 0.9667 & 1.0000 & 1.0000 & 111.31 & 0.85 & 0.84 & 0.9500 & 0.9670 \\
& prun-avg  & 0.7334 & 1.0000 & 1.0000 & 107.13 & 0.78 & 0.80 & 0.7820 & 1.0000 \\
& svd-qkd   & 0.8667 & 1.0000 & 1.0000 & 104.67 & 0.85 & 0.86 & 0.9100 & 0.7340 \\
& qkd       & 0.6667 & 0.6667 & 0.6667 & 104.92 & 0.77 & 0.83 & 0.7420 & 0.9010 \\
\midrule
\multirow{5}{*}{Genomic}
& - & 0.6300 & 0.6333 & 0.7000 & 826.78 & 0.63 & 0.64 & 0.7100 & 0.6300 \\
& prun & 0.6100 & 0.6000 & 0.6333 & 824.67 & 0.63 & 0.63 & 0.6800 & 0.6300 \\
& prun-avg & 0.6000 & 0.5667 & 0.6333 & 821.92 & 0.63 & 0.63 & 0.7000 & 0.6000 \\
& svd-qkd & 0.5933 & 0.6667 & 0.7000 & 830.58 & 0.63 & 0.63 & 0.6900 & 0.6300 \\
& qkd & 0.6600 & 0.6000 & 0.7000 & 838.86 & 0.63 & 0.62 & 0.7000 & 0.6300 \\
\bottomrule
\end{tabular}
}
\label{tab:iris-comprehensive}
\end{table}

\subsubsection{Data Condensation}
\begin{figure}[!htbp]
    \centering
    \begin{subfigure}[b]{0.3\columnwidth}
        \centering
       \includegraphics[width=\columnwidth]{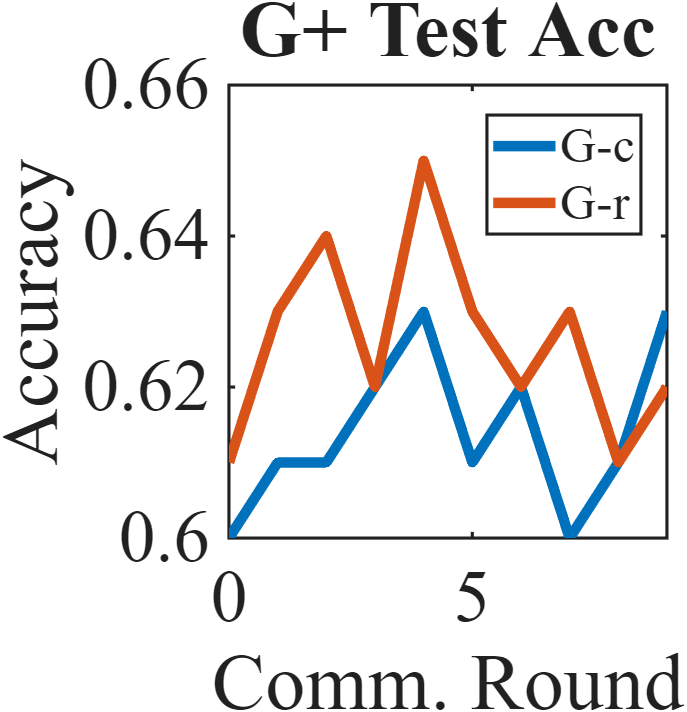}
    \caption{Test Acc - G}
    \label{fig:test_acc_condensation_g}
    \end{subfigure}
    \begin{subfigure}[b]{0.3\columnwidth}
        \centering
       \includegraphics[width=\columnwidth]{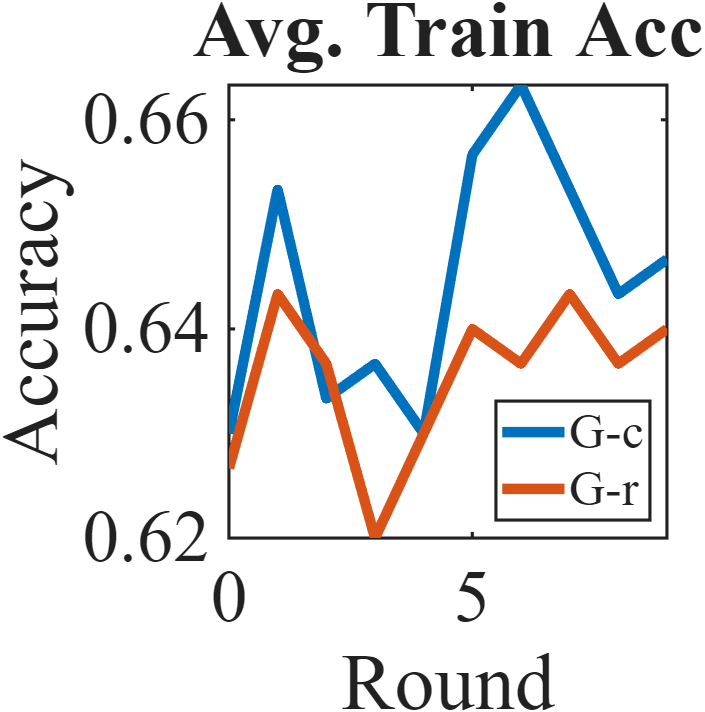}
    \caption{Avg. Train - G}
    \label{fig:train_acc_condensation_g}
    \end{subfigure}
    \begin{subfigure}[b]{0.3\columnwidth}
        \centering
       \includegraphics[width=\columnwidth]{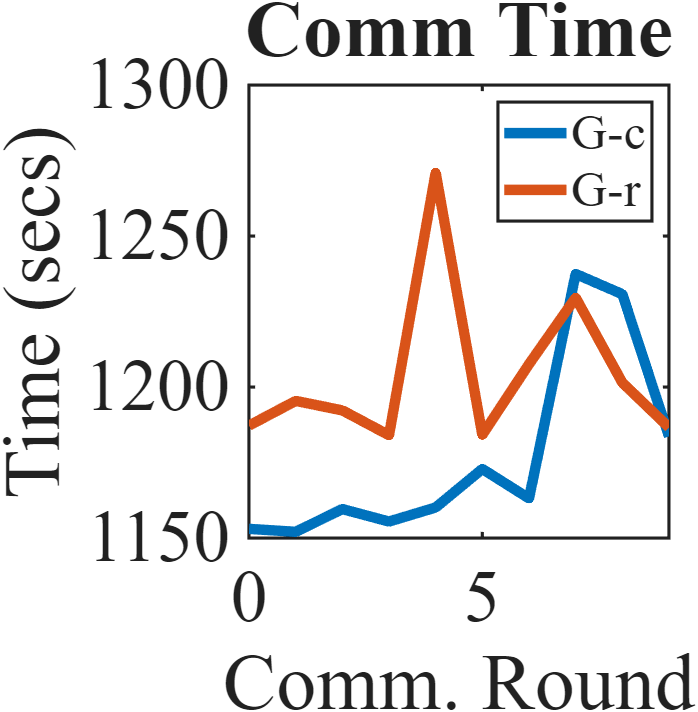}
    \caption{Comm Time - G}
    \label{fig:comm_time_condensation_g}
    \end{subfigure}
     \begin{subfigure}[b]{0.3\columnwidth}
        \centering
       \includegraphics[width=\columnwidth]{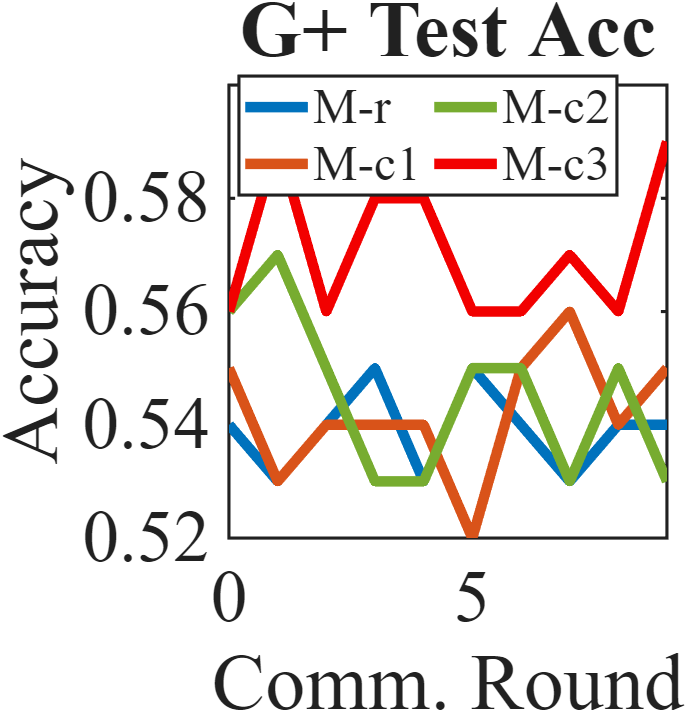}
    \caption{Test Acc - M}
    \label{fig:test_acc_condensation_m}
    \end{subfigure}
    \begin{subfigure}[b]{0.3\columnwidth}
        \centering
       \includegraphics[width=\columnwidth]{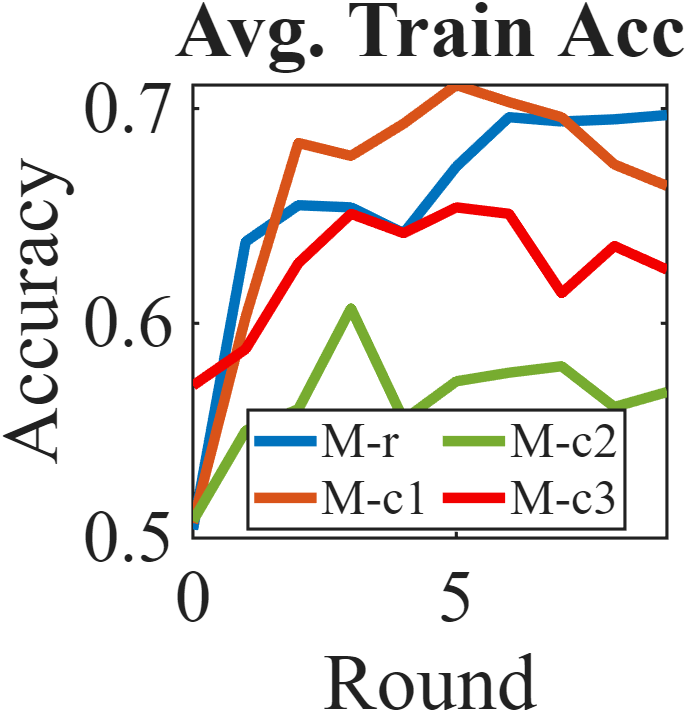}
    \caption{Avg. Train - M}
    \label{fig:train_acc_condensation_m}
    \end{subfigure}
    \begin{subfigure}[b]{0.3\columnwidth}
        \centering
       \includegraphics[width=\columnwidth]{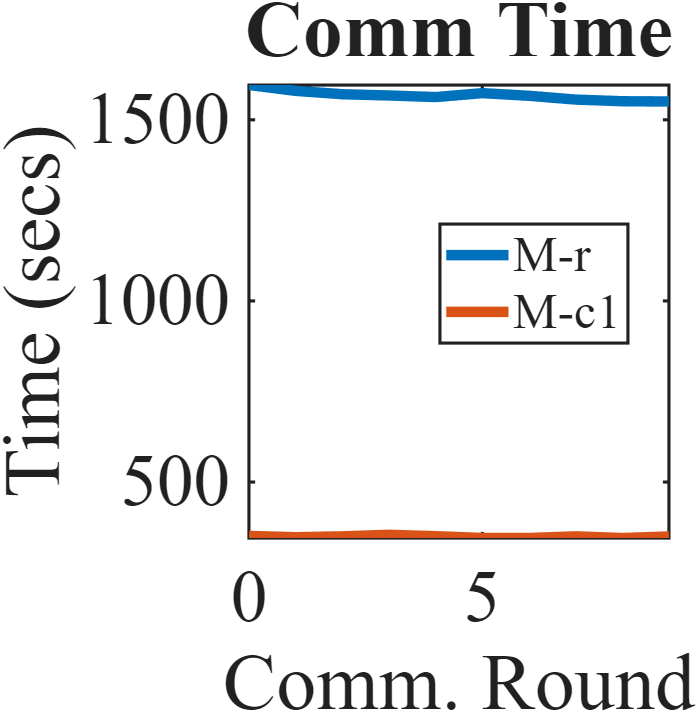}
    \caption{Comm Time - M}
    \label{fig:comm_time_condensation_m}
    \end{subfigure}
    \caption{Performance comparison: Data Condensation with Genomic Dataset (10,000 original samples, 400 condensed samples) and MNIST dataset (18,623 original samples, 600 condensed samples [0, 1, 2 digits]) }
    \label{fig:results_condensation}
\end{figure}

\begin{figure}[!htbp]
    \centering
    \begin{subfigure}[b]{0.3\columnwidth}
        \centering
        \includegraphics[width=\textwidth]{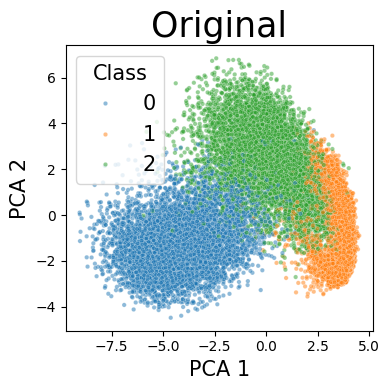}
        \caption{O-MNIST}
        \label{fig:mnist_dataset_original1}
    \end{subfigure}
    \begin{subfigure}[b]{0.3\columnwidth}
        \centering
        \includegraphics[width=\textwidth]{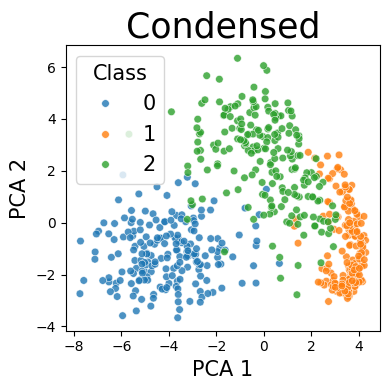}
        \caption{C-MNIST}
        \label{fig:mnist_dataset_condensed1}
    \end{subfigure}
    \begin{subfigure}[b]{0.3\columnwidth}
        \centering
        \begin{subfigure}[b]{\columnwidth}
            \centering
            \includegraphics[width=0.5\textwidth]{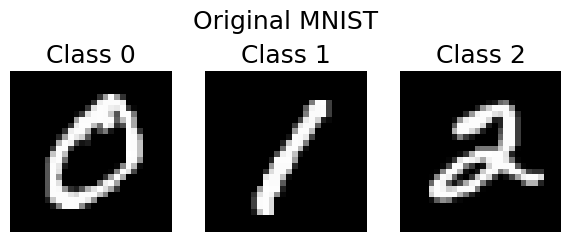}
            \caption{O-Sample}
            \label{fig:mnist_dataset_digits_original}
        \end{subfigure}
        \begin{subfigure}[b]{\columnwidth}
            \centering
            \includegraphics[width=0.5\textwidth]{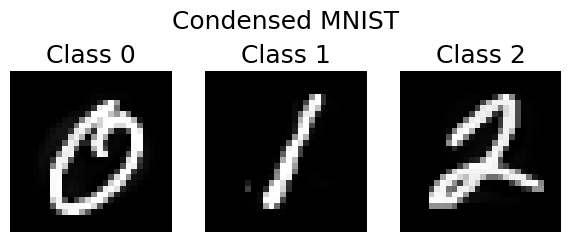}
            \caption{C-Sample}
            \label{fig:mnist_dataset_digits_condensed}
        \end{subfigure}
        \vspace{-0.5em}
        \caption{Sample MNIST}
        \label{fig:mnist_sample_comparison}
    \end{subfigure}
    \begin{subfigure}[b]{0.3\columnwidth}
        \centering
        \includegraphics[width=\textwidth]{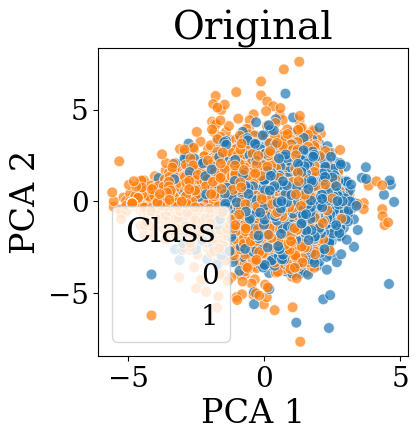}
        \caption{O-Genomic}
        \label{fig:genomic_dataset_original_20000}
    \end{subfigure}
    \begin{subfigure}[b]{0.3\columnwidth}
        \centering
        \includegraphics[width=\textwidth]{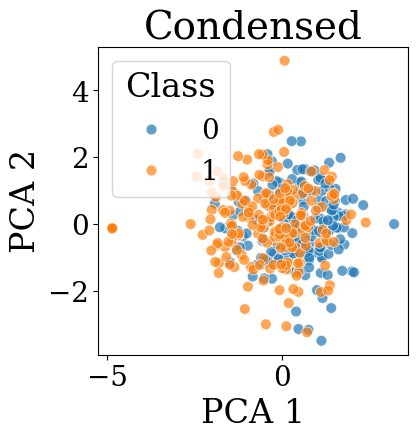}
        \caption{C-Genomic}
        \label{fig:genomic_dataset_condensed_400}
    \end{subfigure}
    \caption{Condensed dataset - C, Original Dataset - O; MNIST and Genomic Data}
    \label{fig:dataset_condensation}
\end{figure}

In terms of data condensation, we generate small synthetic data from the original dataset using Genomic dataset (Figures \ref{fig:genomic_dataset_original_20000}, \ref{fig:genomic_dataset_condensed_400}) and MNIST datasets (Figures \ref{fig:mnist_dataset_digits_original}, \ref{fig:mnist_dataset_digits_condensed}, \ref{fig:mnist_sample_comparison}).
In Figure \ref{fig:results_condensation}, we observe that condensation definitely has some impact on test accuracy as seen in Figure \ref{fig:test_acc_condensation_g} with a condensed dataset performing better for the genomic dataset (-G).
In terms of communication time, as in Figure \ref{fig:comm_time_condensation_g}, the impact is clear, as we can observe a drastic reduction in communication overhead.
This result is promising, as this is achieved in addition to the lack of performance degradation seen in Figure \ref{fig:test_acc_condensation_g}.
If we were to compare the varying size of the data set itself, as in Figure \ref{fig:comm_time_condensation_g}, 
the impact of data condensation is clear, with more advantages in terms of communication overhead.
However, this experiment was done with only 3 device dataset condensed from 20000 samples to 400, thus communication would hugely be impacted if we were to use full dataset and more devices.
 Using the MNIST dataset, as in Figures \ref{fig:test_acc_condensation_m}, \ref{fig:train_acc_condensation_m}, and \ref{fig:comm_time_condensation_m}, 
 we observe that with the G+ test accuracy and the average train accuracy of the devices, the results are comparable, whereby in one instance even perform better (M-c2, Figure \ref{fig:train_acc_condensation_m}) than with the real set (M-r) and in communication time, with real dataset as it is bigger in size causes more communication bottleneck.

\section{Conclusion}
In this work, we introduced a privacy-preserving quantum federated 
learning (QFL) framework. We examined a range of protocols and methodologies from standard QFL to perform several ablation studies. 
Furthermore, we provided an extensive theoretical analysis of different 
approaches and demonstrated how they can be integrated into the QFL 
framework to strengthen privacy guarantee within QFL. 
In future work, we plan to develop quantum privacy protocols and investigate more advanced privacy-preserving frameworks for QFL.


\end{document}